\documentclass[aps,prd,preprintnumbers,showpacs,superscriptaddress,nofootinbib,amsmath,amssymb,floats,floatfix,showkeys,notitlepage,longbibliography]{revtex4-1}
\usepackage{comment}
\usepackage{graphicx}
\usepackage{subfigure}
\usepackage{palatino}
\usepackage[commandnameprefix=always]{changes}
\usepackage{hyperref}
\hypersetup{colorlinks=true,linkcolor=red,urlcolor=red,citecolor=red}
\usepackage[toc,page]{appendix}
\usepackage[normalem]{ulem}

\usepackage{lipsum}
\usepackage{graphicx}
\usepackage{subfigure}
\usepackage{palatino}
\usepackage{float}
\usepackage{sans}
\usepackage{adjustbox}
\usepackage{latexsym}
\usepackage{amsmath}
\usepackage{amssymb}
\usepackage{amsfonts}
\usepackage{dcolumn}
\usepackage{bm}
\usepackage{tikz}
\usepackage{bigints}
\usepackage{array,tabularx,multirow,booktabs}
\usepackage[tracking=true]{microtype}
\SetTracking{}{500}
\SetTracking{encoding={*}, shape=sc}{40}
\allowdisplaybreaks
\usepackage{adjustbox}
\usepackage{latexsym}
\usepackage{amsmath}
\usepackage{amssymb}
\usepackage{amsfonts}
\usepackage{dcolumn}
\usepackage{bm}
\usepackage{tikz}
\usepackage{bigints}
\usepackage{array,tabularx,multirow,booktabs}
\usepackage[tracking=true]{microtype}
\usepackage{color}
\allowdisplaybreaks

\begin{document}

\title{ WGC as WCCC protector: The Synergistic Effects of various Parameters in Identifying WGC candidate Models}

\author{Mohammad Ali S. Afshar}
\email{m.a.s.afshar@gmail.com}
\affiliation{Department of Physics, Faculty of Basic Sciences,
University of Mazandaran\\
P. O. Box 47416-95447, Babolsar, Iran}

\author{Jafar Sadeghi}
\email{pouriya@ipm.ir}
\affiliation{Department of Physics, Faculty of Basic Sciences,
University of Mazandaran\\
P. O. Box 47416-95447, Babolsar, Iran}

\begin{abstract}
The integration of non-commutative geometry and Gauss-Bonnet corrections in an action and the study of their black hole responses can provide highly intriguing insights. Our primary motivation for this study is to understand the interplay of these two parameters on the geodesics of spacetime, including photon spheres and time-like orbits.
In this study, we found that this integration, in its initial form, can limit the value of the Gauss-Bonnet parameter ($\alpha$), creating a critical threshold beyond which changes in the non-commutative parameter ($\Xi$) become ineffective, and the structure can only manifest as a naked singularity.
Furthermore, we found that using a more complex model, which includes additional factors such as a cloud of strings and linear charge, as a sample for studying spacetime geodesics, yield different and varied results.
In this scenario, negative $\alpha$ values can also play a role, notably preserving the black hole form even with a super-extremal charge ($q > m$). For $\alpha> 0.1$, the black hole mass parameter becomes significantly influential, with a critical mass below which the impact of other parameter changes is nullified. Interestingly, considering a more massive black hole, this high-mass state also maintains its black hole form within the super-extremal charge range.
The existence of these two models led us to our main goal. By examining the temperature for these two cases, we find that both situations are suitable for studying the Weak Gravity Conjecture (WGC). Finally, based on the behavior of these two models, we will explain how the WGC acts as a logical solution and a protector for the WCCC.
\end{abstract}

\date{\today}

\maketitle
\section{Introduction}
In the previous article \cite{1}, we discussed that due to the theoretical and practical limitations of current human  knowledge, it is currently impossible to achieve a unified and comprehensive model for the structure of black holes.  Therefore, various theoretical models are constructed and examined based on the combination of existing and known fields, with the aim of selecting the model that best aligns with nature and empirical data in the future \cite{2,3,4,5,6,7,8,9,10}. One of these models utilizes the properties of Non-Commutativity(NC), means the space-time coordinates do not commute, to study the geometry of spacetime. This model, while attempting to resolve the central singularity, also impacts the physical properties of black holes such as the event horizon and mass distribution. Our first goal in this series of articles is to examine these impact on the structure of  geodesics and circular orbits of black holes influenced by NC. In previous work \cite{1}, we explored models resulting from  the combination of gravity and electromagnetism in both linear and nonlinear forms. We observed that in the linear form of charge, despite the mass distribution changing from a point to a Gaussian distribution, the structural behavior of  geodesics, including photon spheres and Time-like Circular Orbits(TCOs), remained very close to the Reissner-Nordström form. However, this combination imposed a constraint on the NC parameter, $\Xi$, such that, for example, given the selected parameter values, the structure retained its black hole form only within the range $0 < \Xi \leq 0.2184$.
In the nonlinear form, we used a model incorporated the Born-Infeld field and added the cosmological constant to the model. Since smaller values of $\Xi$ result in better alignment with reality, We found that the inclusion of the nonlinear field significantly improved the model’s accuracy. Also, we observed that, in the dS model, the curvature constant played a much more critical role, creating a critical curvature constant beyond which the structure would no longer exhibit black hole behavior.\\
The primary objective of this series of articles is to develop models that can maintain the necessary initial conditions for black hole formation in both extremal and super-extremal charge regimes. Specifically, these models should simultaneously exhibit an event horizon and an unstable photon sphere. The significance of super-extremal black holes lies in the increased likelihood of the existence of ultra-extremal particles. This not only strengthens the Weak Gravity Conjecture (WGC) but also, as we will demonstrate, reinforces the Weak Cosmic Censorship Conjecture (WCCC).
In our previous study, we found that only the Born-Infeld model exhibits such capacity [1].\\
Based on the above experiences, in this article, we will first look at structures that are mainly related to the concepts of gravity and curvature, and examine them. During the 1970s and 1980s, theorists began investigating extensions of General Relativity, including higher-order curvature terms, to explain phenomena such as singularities or high-energy gravitational waves that were not fully accounted for by General Relativity. The Gauss-Bonnet theorem connects geometry and topology, specifically in the context of curved surfaces. In simple terms, the theorem relates the total curvature of a surface to its topological characteristics, such as the number of holes or the genus of the surface \cite{11,12}. Einstein-Gauss-Bonnet gravity extends General Relativity by incorporating higher-order curvature terms, particularly the Gauss-Bonnet term\cite{13}. Therefore, in the first step, we will examine the Gauss-Bonnet structure to assess the impact of non-commutativity on this model, especially on the geodesic structure.\\
Subsequently, we will consider the combination of this model with a linear charge and the addition of a string cloud. A cloud of strings refers to a theoretical construct where a collection of one-dimensional strings (from string theory) are distributed in space. This concept is analogous to a cloud of dust but with strings instead of particles. These strings can interact with gravitational fields and other forces in the universe \cite{14}. When a black hole is surrounded by a cloud of strings, it can significantly affect the black hole's properties and behavior. The presence of a string cloud can alter the spacetime geometry around the black hole \cite{15,16}. Also, strings in the cloud can interact with particles near the black hole, affecting their trajectories and energy dynamics. This can lead to unique observational signatures that help scientists study the fundamental properties of black holes and string theory \cite{17}.\\Finally, we will revisit the concept of the WGC. After examining whether these new models can exhibit a state of super-extremality, we will explore the relationship between temperature, the super-extremal model, and the WGC. We will investigate the conditions under which the models under study can provide further evidence to satisfy the WGC from a thermal perspective.Based on the above information, we will organize this paper as follows:\\
In Section 2, since in the previous work \cite{1} we have sufficiently covered the principles and foundations, we will refer to that and only highlight the main equations. In the next two sections, we will study different models using the mathematical method introduced. In Section 5, we examine extreme and superextreme conditions and show that adding the temperature condition to the models under consideration can be a complement to make these definitions more precise, and accordingly, we can better understand the conditions that will lead to the realization of WGC. Finally we will present the results of our analysis in Section 6.
\section{Methodology}
As mentioned in the introduction, since there is a detailed explanation and development in the previous work \cite{1} and references \cite{18,19,20,21}, we will present the principles and foundations only in the form of highlighting practical relationships in the following two sections.
\subsection{Topological Photon Sphere }
To study the photon sphere, we use the topological method instead of the traditional method, a different method that, due to the general characteristics of its mathematical structure \cite{21.1,21.2}, has been extended not only to the study of light rings but also to thermodynamics\cite{21.3,21.4,21.5,21.6,21.7,21.8,21.9,21.10}.\\
In order to use the concept of mapping and topological charge, we need a vector field. If $\varphi_{r}$  and $\varphi_{\theta}$ are components of this general vector field, that is:\cite{18}
\begin{equation}\label{1}
\varphi=(\varphi_{r}, \varphi_{\theta}),
\end{equation}
 also here, we can  rewrite the vector as $\varphi=||\varphi||e^{i\theta}$, where $||\varphi||=\sqrt{\varphi_{j}\varphi_{j}}$, or $\varphi = \varphi_{r} + i\varphi_{\theta}$.
 Based on this, the normalized vector is defined as,
 \begin{equation}\label{2}
n^j=\frac{\varphi_{j}}{||\varphi||},
\end{equation}
 where $j=1,2$  and  $(\varphi_{1}=\varphi_{r})$ , $(\varphi_{2}=\varphi_{\theta})$.
In the next stage, with respect to the necessity of spherical symmetry as a prerequisite for studying this method, and given the most general form of the metric in 4-dimensional form we have:
\begin{equation}\label{(3)}
\mathit{ds}^{2}=-\mathit{dt}^{2} f \! \left(r \right)+\frac{\mathit{dr}^{2}}{g \! \left(r \right)}+\left(d\theta^{2}+d\phi^{2}
\sin \! \left(\theta \right)^{2}\right) h \! \left(r \right)=\mathit{dr}^{2} g_{\mathit{rr}}-\mathit{dt}^{2} g_{\mathit{tt}}+d\theta^{2} g_{\theta \theta}+d\phi^{2} g_{\phi \phi},
\end{equation}
for the new form of effective potential,we have \cite{18}:
\begin{equation}\label{(4)}
\begin{split}
\mathbb{H}(r, \theta)=\sqrt{\frac{-g_{tt}}{g_{\phi\phi}}}=\frac{1}{\sin\theta}\bigg(\frac{f(r)}{h(r)}\bigg)^{1/2}.
\end{split}
\end{equation}
With respect to this potential, now the vector field components become as follows:
\begin{equation}\label{(5)}
\begin{split}
&\varphi_{r}=\frac{\partial_r\mathbb{H}}{\sqrt{g_{rr}}}=\sqrt{g(r)}\partial_{r}\mathbb{H},\\
&\varphi_{\theta}=\frac{\partial_\theta \mathbb{H}}{\sqrt{g_{\theta\theta}}}=\frac{\partial_\theta \mathbb{H}}{\sqrt{h(r)}}.
\end{split}
\end{equation}
Now, using the above relations and considering the definition of topological charges, for each photon sphere we can assign a charge. To avoid repetition, a detailed description of how to calculate it is available in references \cite{1,18,19,20,21}.
\subsection{Time-like Circular Orbits (TCOs) }
Although the geometry and dynamics of spacetime within the event horizon seem beyond our reach, beyond the event horizon, our field equations retain their logical applicability. This allows us to precisely identify and classify geometric-kinematic boundaries. Accordingly, we can consider the following classifications:\\
1. Light-like Geodesics : Massless particles and photons trajectories.\\
2. Timelike Geodesics: Massive objects trajectories .\\
Both types of theses geodesics, depending on the characteristics of the gravitational potential, may either lack observable turning points or possess a turning point, forming loops. These loops can be stable or unstable. In the case of null geodesics, these loops are referred to as photon spheres, while in timelike geodesics, they are known as Timelike Circular Orbits (TCOs).\\
To better understand the behavior of these orbits and avoid mathematical complexities, we make certain assumptions that do not compromise the generality of our analysis. For instance, in this study, we consider a static and axisymmetric spacetime with $\mathbb{Z}_2$ symmetry in a 1+3 dimensional framework. Based on the metric equation Eq. (\ref{(3)}), we consider the following quantities \cite{22}:
\begin{equation}\label{(6)}
\mathbb{A} =g_{\phi \phi} E^{2}+g_{\mathit{tt}} L^{2},
\end{equation}
\begin{equation}\label{(7)}
\mathbb{B} =-g_{\phi \phi} g_{\mathit{tt}},
\end{equation}
where the energy and angular momentum are denoted by E, L. Now the Lagrangian can be recast as:
\begin{equation}\label{(8)}
2 \mathfrak{L} =-\frac{\mathbb{A}}{\mathbb{B}}=\varrho,
\end{equation}
where $\varrho = -1, 0 $ for time-like, light-like  geodesics, respectively. With the above Lagrangian, the effective potential can be rewritten as follows:
\begin{equation}\label{(9)}
\mathbb{V}_{\mathit{eff} \! \left(\varrho \right)}\! \left(r \right)=\varrho +\frac{\mathbb{A}}{\mathbb{B}}.
\end{equation}
For an observer at infinity, the angular velocity in terms of metric parameters will be as follows::
\begin{equation}\label{(10)}
\mathbf{\Omega}_{\pm}=\frac{g_{\mathit{tt}} L}{g_{\phi \phi} E},
\end{equation}
we also consider the $\beta$ quantity in the following form:
\begin{equation}\label{(11)}
\beta_{\pm}=-\mathbb{A} \! \left(r^{\mathit{cir}},\mathbf{\Omega}_{\pm},\mathbf{\Omega}_{\pm}\right),
\end{equation}
in which $\pm$ is a sign of prograde/retrograde orbits \cite{22}.
\section{Non-Commutative black hole in 4D Einstein-Gauss-Bonnet theory}
\begin{figure}[H]
 \begin{center}
 \subfigure[]{
 \includegraphics[height=5cm,width=5cm]{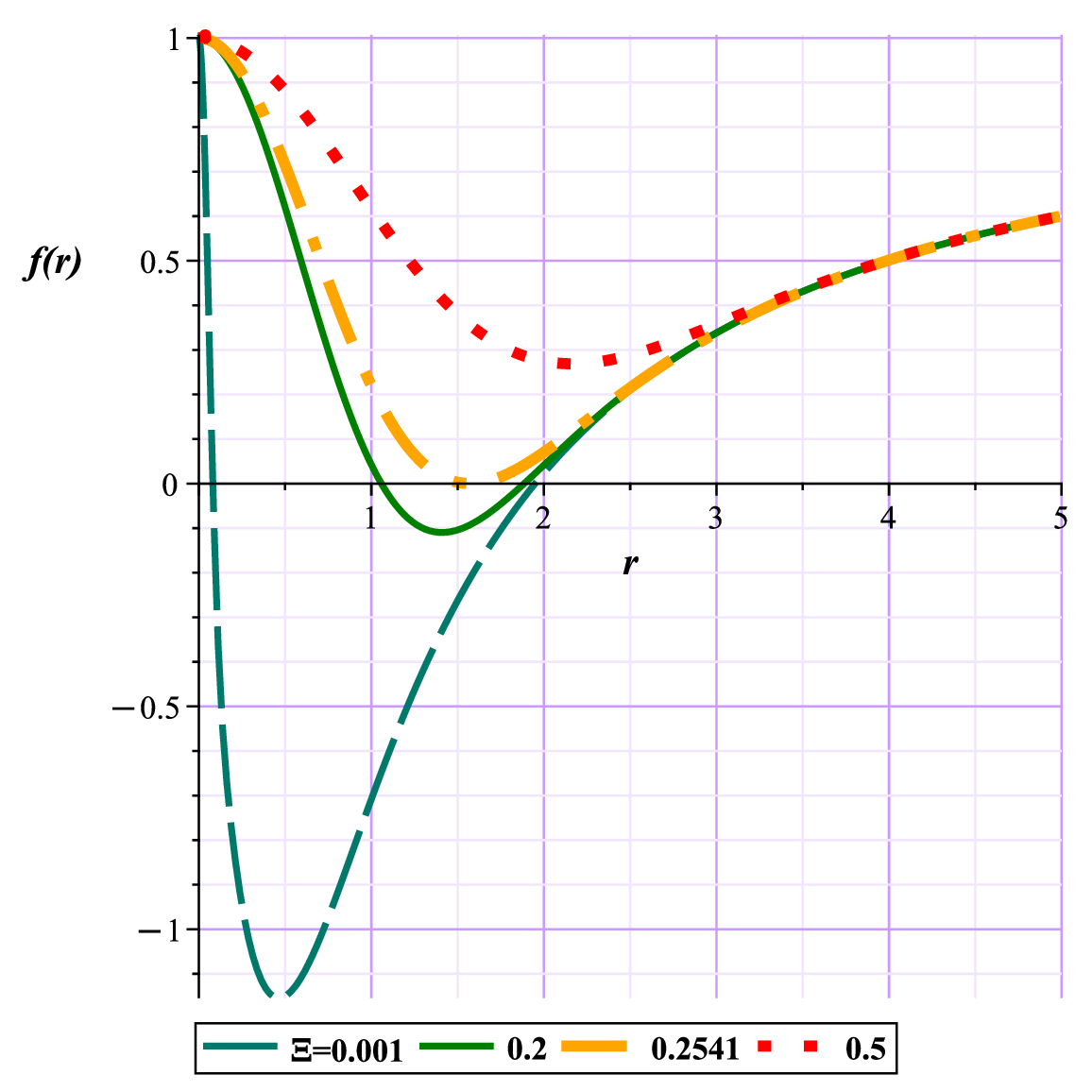}
 \label{1a}}
 \subfigure[]{
 \includegraphics[height=5cm,width=5cm]{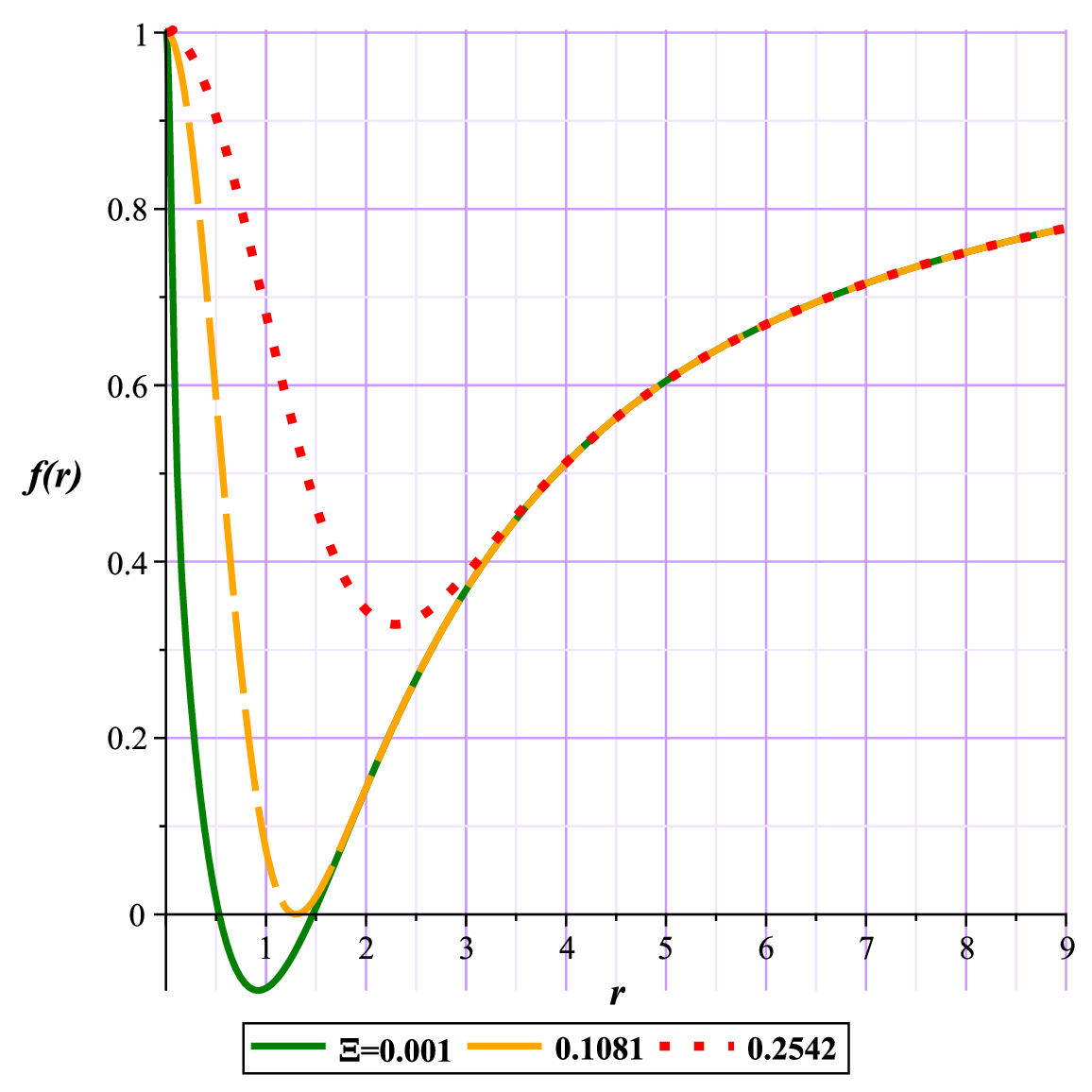}
 \label{1b}}
 \subfigure[]{
 \includegraphics[height=5cm,width=5cm]{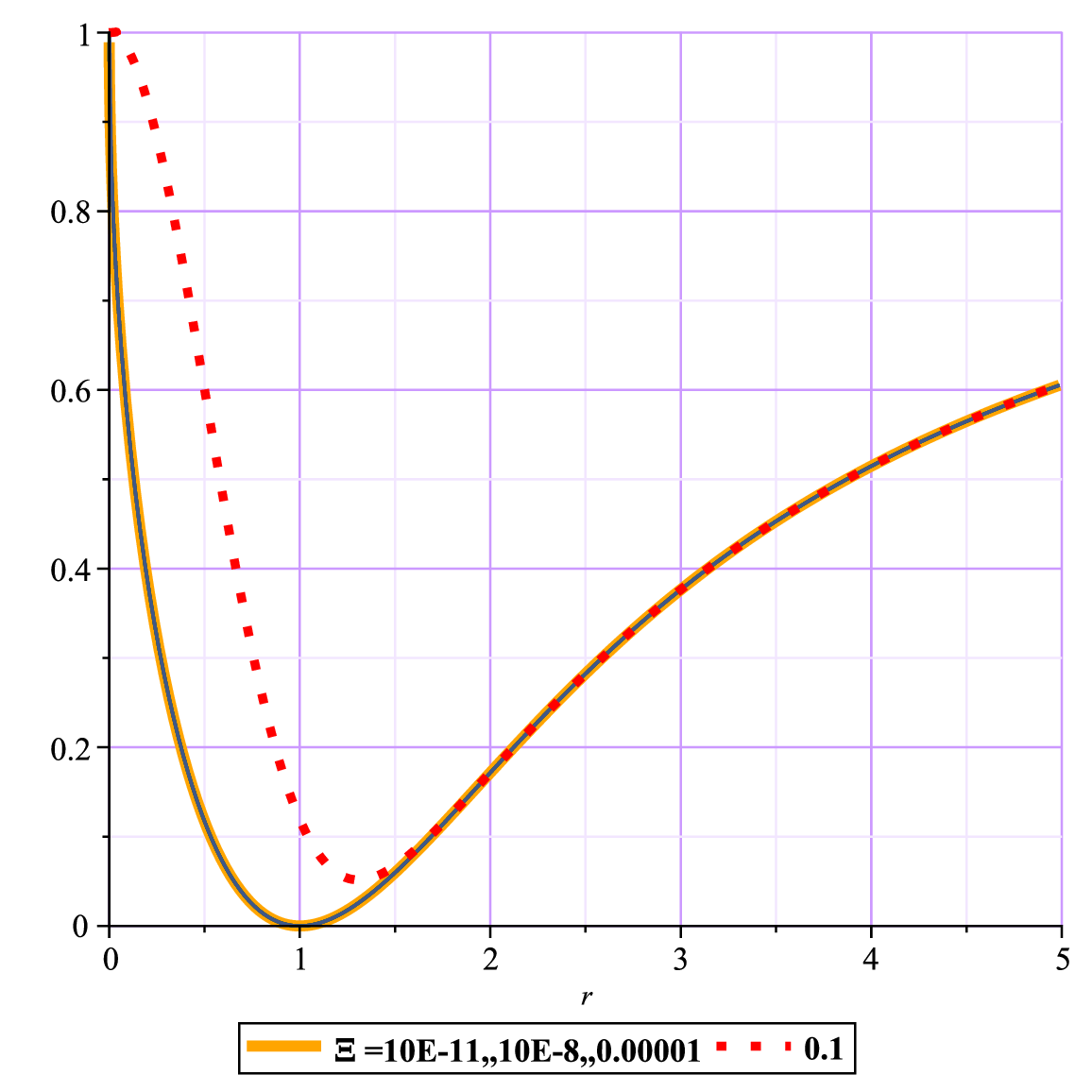}
 \label{1c}}
 \caption{\small{Metric function with different $\Xi$ for  NCEGB BH, Fig (1a) $\alpha=0.05$, Fig (1b) $\alpha=0.39$, Fig (1c) $\alpha=0.5$  }}
 \label{1}
\end{center}
\end{figure}

The Einstein-Gauss-Bonnet (EGB) black hole arises from the Einstein-Gauss-Bonnet gravity, which is an extension of General Relativity  incorporating higher-order curvature corrections. This theory includes the Gauss-Bonnet term, which is a specific combination of curvature invariants that appears naturally in the low-energy limit of string theory. The Gauss-Bonnet term is significant in higher-dimensional spacetimes (more than four dimensions) and was first introduced by David Lovelock in 1971 \cite{23}. EGB gravity provides solutions in higher-dimensional spacetimes, which are essential in string theory and other advanced theoretical frameworks \cite{23}. Some solutions in EGB gravity can avoid the singularities that are typically present in black hole solutions of GR, leading to more physically realistic models \cite{24}. The metric for the black hole that we consider is \cite{25}:
\begin{equation}\label{(12)}
f(r)=1+\frac{\left(1-\sqrt{1+\frac{32 \gamma \left(\frac{3}{2},\frac{r^{2}}{4 \Xi}\right) m \alpha}{\sqrt{\pi}\, r^{3}}}\right) r^{2}}{4 \alpha}.
\end{equation}
The NC parameter is $\Xi$, the total mass of the source is m, $\alpha$ is a coupling constant with dimension of $(length)^{2}$ which is positive in heterotic string theory and the lower incomplete Gamma function is represented with $\gamma$, which can be written in the following form in terms of $\Gamma$ (the upper incomplete Gamma function) :
\begin{equation}\label{(13)}
 \gamma (\frac{3}{2},\frac{r^{2}}{4\,\Xi})\equiv \int_{0}^{\frac{r^{2}}{4\,\Xi}}\sqrt{t}\times {\mathrm e}^{t}\quad d t =\Gamma (\frac{3}{2})-\Gamma (\frac{3}{2},\frac{r^{2}}{4\,\Xi}),
\end{equation}
As previously stated, the smearing of the mass distribution in Non-Commutative Black Holes (NCBH),\\ 
-mass is spread over a finite region against being at a single point-
impacts the structure of the black hole’s event horizon and, consequently, parameter $\Xi$ playing a significant role in determining the event horizon.
An important point to mention before any calculations is that in all the models studied in this work, due to the unsolvability of the equations in their general form, we are compelled to solve them numerically. In the Gauss-Bonnet model, as clearly illustrated in Fig.(\ref{1a}), the behavior of the metric function still depends on the value of $\Xi$.
Our study shows that, for instance, for $ m = 1$ and at very small values of $ \alpha$ (e.g., $ \alpha = 0.05$), the permissible range for $\Xi$ is $ 0 < \Xi < 0.2541$, indicating that the behavior of the model closely resembles the Schwarzschild form ($ 0 < \Xi_{Sch} < 0.2758 $ in Appendix 1). As alpha increases and the range of $\Xi$ gradually decreases, the model appears to approach physical reality more closely. For example, for $ \alpha = 0.39 $, the range of $\Xi$ is reduced to $0 < \Xi < 0.1081$, which is a relatively significant reduction, as shown in Fig.(\ref{1b}).
A particularly interesting aspect of this model is the existence of a critical limit for the $\alpha$ parameter, beyond which the influence of the$ \Xi$ parameter vanishes, and the structure exhibits the behavior of an extremal black hole, Fig.(\ref{1c})).
In Figure (1c), three graphs for different $\Xi$, namely $10^{-11}$, $10^{-8}$ and $10^{-5}$, are superimposed, indicating that the change in the parameter $\Xi$ has no effect.
\subsection{Topological Photon Sphere }
According to the metric function Eq. (\ref{(3)}) and also according to the following equations:
\begin{equation}\label{(14)}
f \! \left(r \right)=g \! \left(r \right),
\end{equation}
\begin{equation}\label{(15)}
h \! \left(r \right)=r,
\end{equation}
and with respect to Eq. (\ref{(4)}), Eq. (\ref{(5)}) we have:
\begin{equation*}\label{(0)}
\begin{split}
h =1+\frac{32 \left(\frac{\sqrt{\pi}}{2}-\frac{\sqrt{\frac{r^{2}}{\Xi}}\, {\mathrm e}^{-\frac{r^{2}}{4 \Xi}}}{2}-\frac{\sqrt{\pi}\, \mathrm{erfc}\left(\frac{\sqrt{\frac{r^{2}}{\Xi}}}{2}\right)}{2}\right) m \alpha}{\sqrt{\pi}\, r^{3}},
\end{split}
\end{equation*}
\begin{equation}\label{(16)}
\begin{split}
\mathbb{H} =\frac{\sqrt{4+\frac{\left(1-\sqrt{h}\right) r^{2}}{\alpha}}}{2 \sin \! \left(\theta \right) r}.
\end{split}
\end{equation}
\begin{equation*}\label{(0)}
\begin{split}
\varphi_{1}=-3 \sqrt{\pi}\, \mathrm{erf}\! \left(\frac{\sqrt{\frac{r^{2}}{\Xi}}}{2}\right) \Xi  m +\frac{\sqrt{\frac{r^{2}}{\Xi}}\, m \left(r^{2}+6 \Xi \right) {\mathrm e}^{-\frac{r^{2}}{4 \Xi}}}{2},
\end{split}
\end{equation*}
\begin{equation*}\label{(0)}
\begin{split}
\varphi_{2}=\pi^{\frac{1}{4}} \sqrt{\frac{16 \sqrt{\pi}\, \mathrm{erf}\! \left(\frac{\sqrt{\frac{r^{2}}{\Xi}}}{2}\right) \alpha  m +\sqrt{\pi}\, r^{3}-16 \sqrt{\frac{r^{2}}{\Xi}}\, {\mathrm e}^{-\frac{r^{2}}{4 \Xi}} \alpha  m}{r^{3}}}\, \Xi  r
\end{split}
\end{equation*}
\begin{equation}\label{(17)}
\begin{split}
\varphi_{r}=-\frac{\left(\varphi_{1}+\varphi_{2}\right) \csc \! \left(\theta \right)}{\pi^{\frac{1}{4}} \sqrt{\frac{16 \sqrt{\pi}\, \mathrm{erf}\left(\frac{\sqrt{\frac{r^{2}}{\Xi}}}{2}\right) \alpha  m +\sqrt{\pi}\, r^{3}-16 \sqrt{\frac{r^{2}}{\Xi}}\, {\mathrm e}^{-\frac{r^{2}}{4 \Xi}} \alpha  m}{r^{3}}}\, r^{3} \Xi}.
\end{split}
\end{equation}
\begin{equation}\label{(18)}
\begin{split}
\varphi_{\theta}=-\frac{\sqrt{4+\frac{\left(1-\sqrt{h}\right) r^{2}}{\alpha}}\, \cos \! \left(\theta \right)}{2 \sin \! \left(\theta \right)^{2} r^{2}}.
\end{split}
\end{equation}
\begin{center}
\textbf{Case I: TTC =-1}
\end{center}
For m=1 and $\alpha=0.39$ we have:
\begin{figure}[H]
 \begin{center}
 \subfigure[]{
 \includegraphics[height=5.5cm,width=6cm]{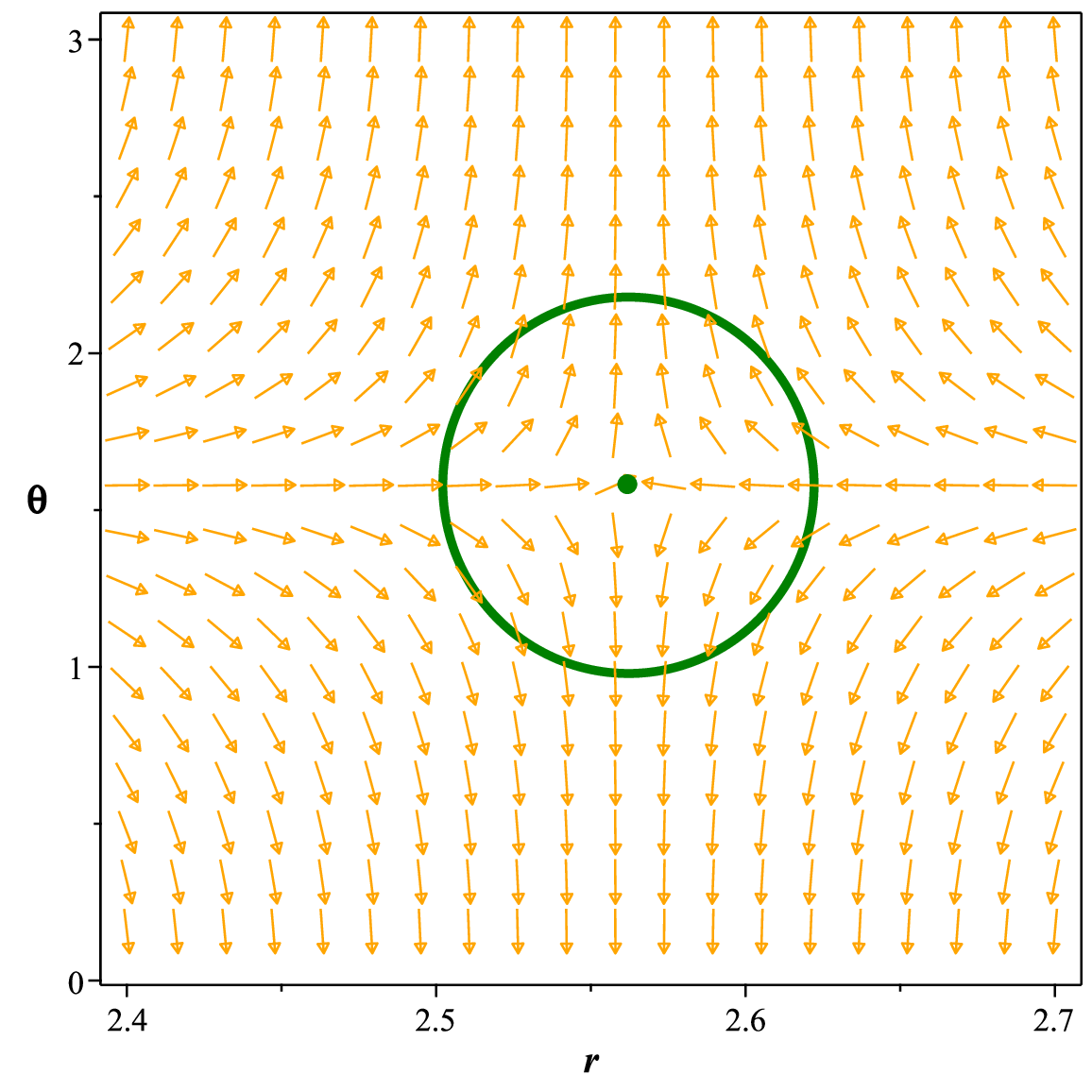}
 \label{2a}}
 \subfigure[]{
 \includegraphics[height=5.5cm,width=6cm]{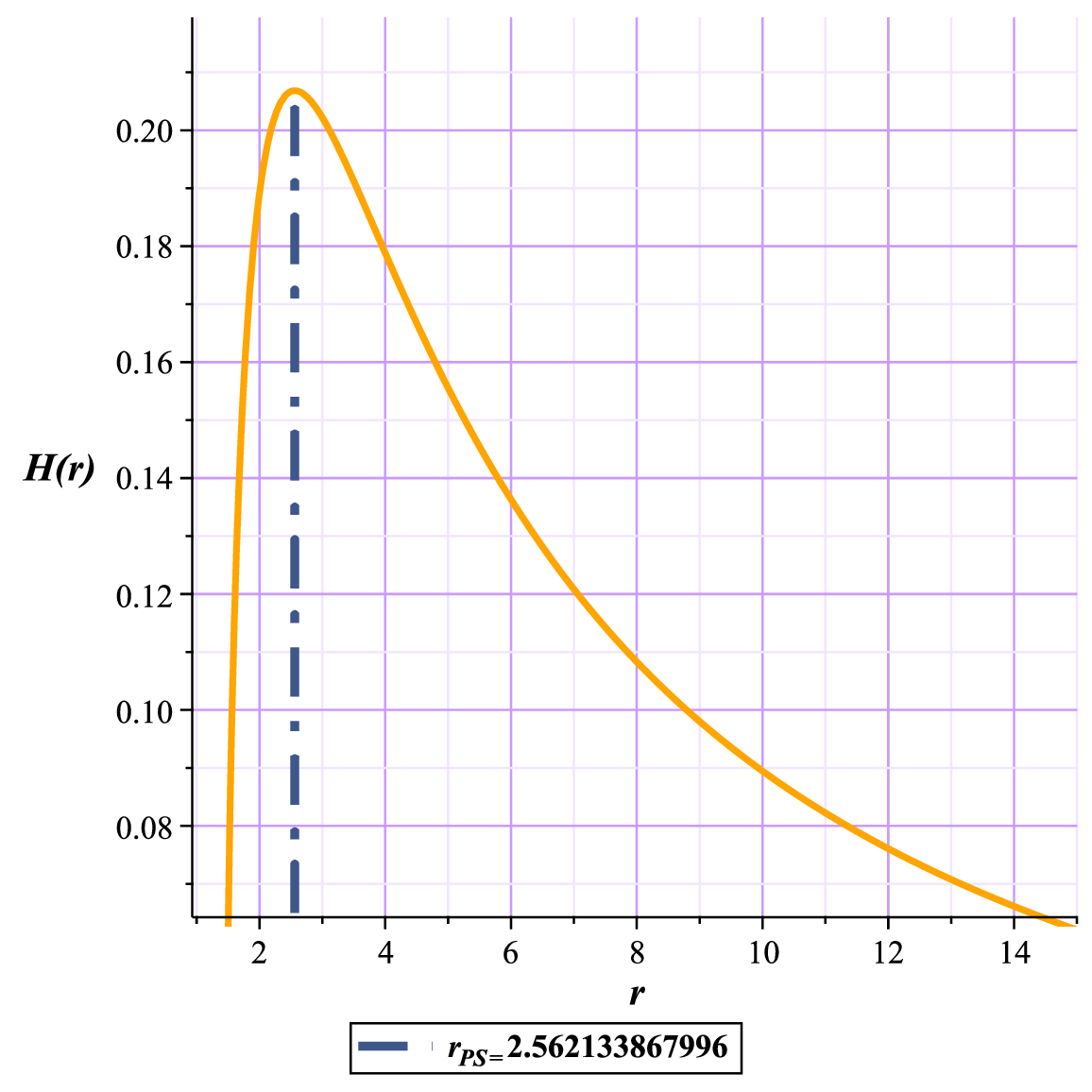}
 \label{2b}}
   \caption{\small{Fig (2a):  The photon sphere location at $ (r, \theta)=(2.736931688316, 1.57)$ with respect to $( \Xi = 0.001, \alpha = 0.39, m = 1 )$ in the $(r-\theta)$ plane of the normal vector field $n$ , (2b): the topological potential H(r) for NCEGB BH  }}
 \label{2}
\end{center}
\end{figure}
In the Fig. (\ref{2a}) we can see that the appearance  photon sphere has a Total Topological Charge (TTC) of -1  and, This, as can be seen in the H diagram, was equivalent to an unstable maximum, Fig. (\ref{2b}). Consequently, this case presents a black hole containing an unstable photon sphere.
\begin{center}
\textbf{Case II: TTC = 0 }
\end{center}
\begin{figure}[H]
 \begin{center}
 \subfigure[]{
 \includegraphics[height=5.5cm,width=6cm]{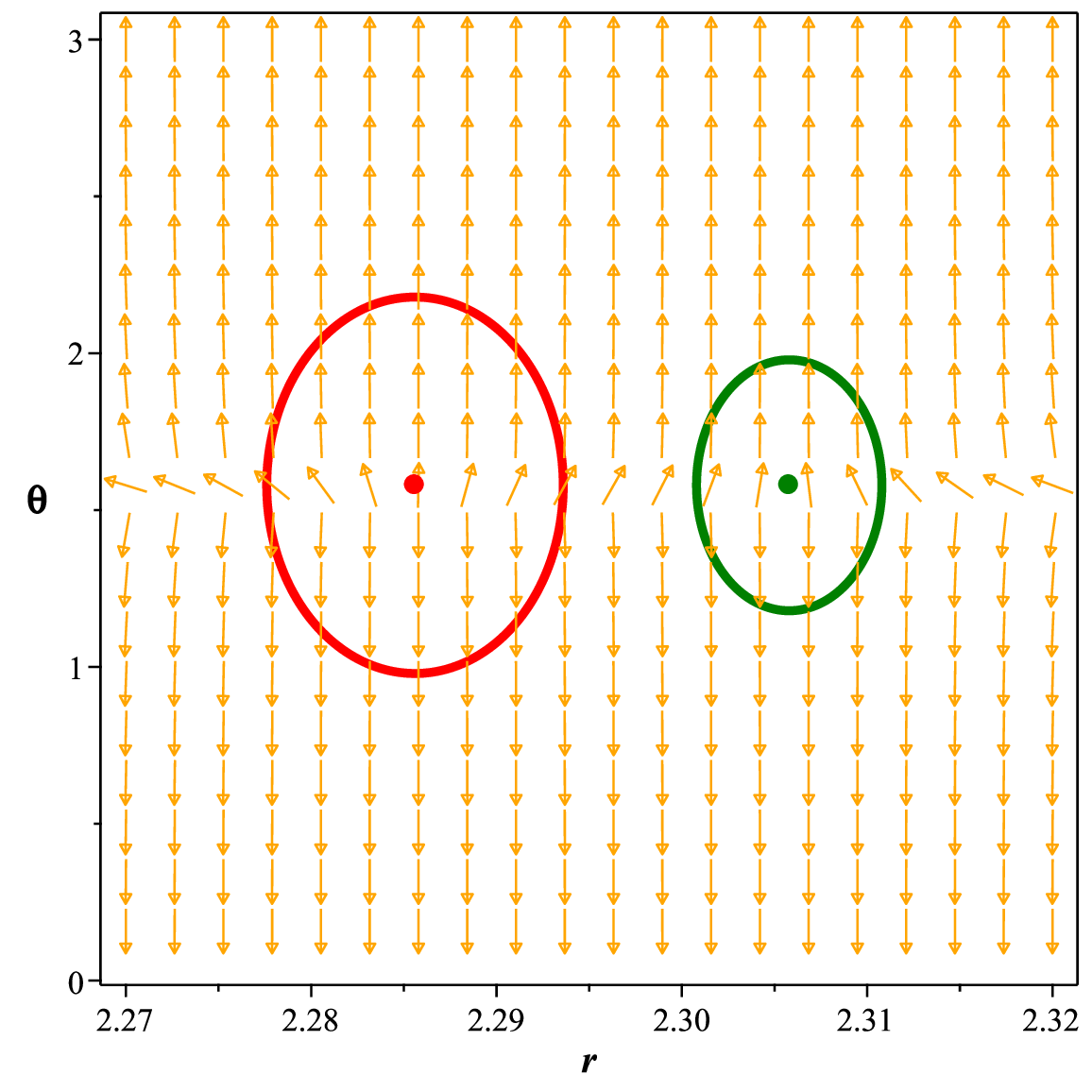}
 \label{3a}}
 \subfigure[]{
 \includegraphics[height=5.5cm,width=6cm]{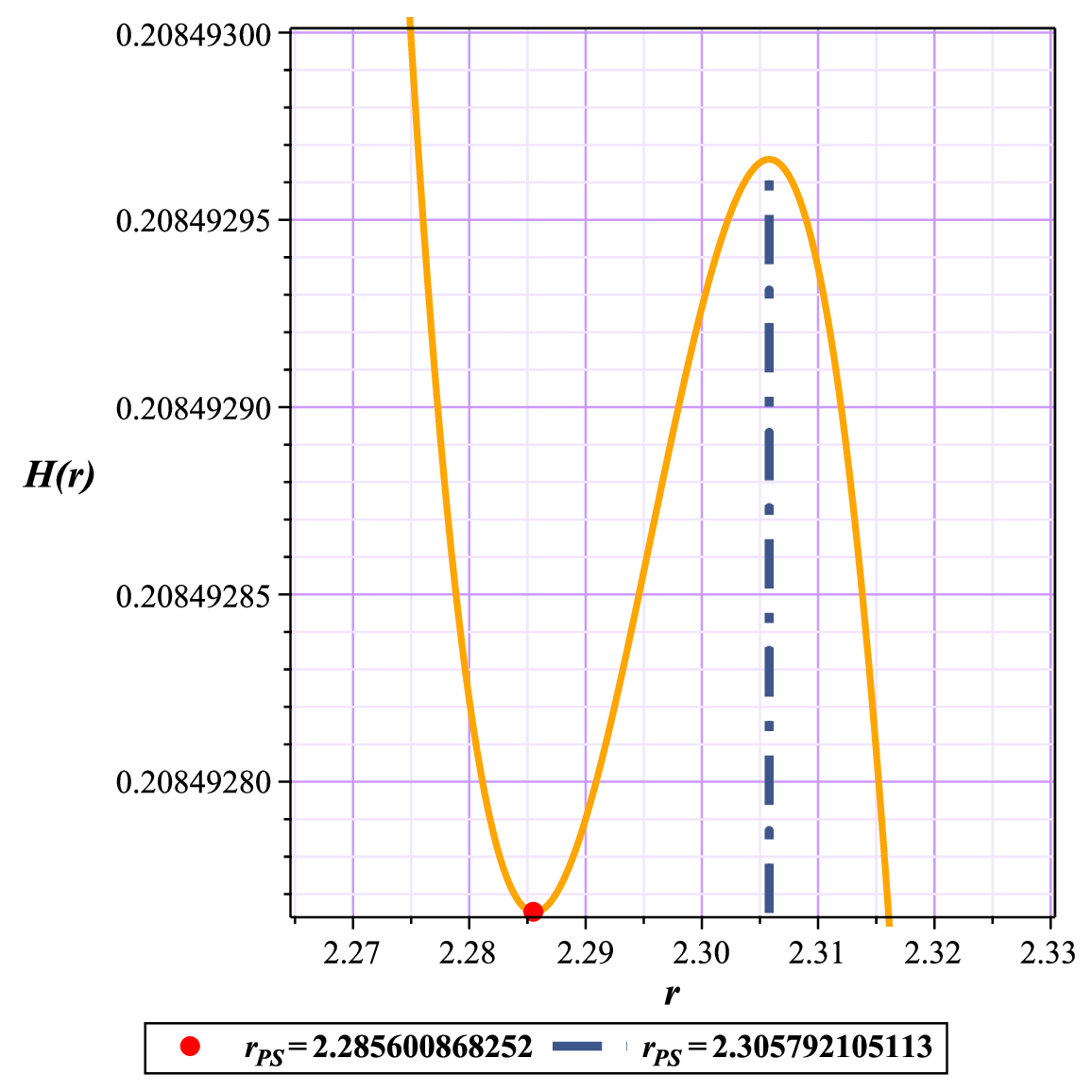}
 \label{3b}}
   \caption{\small{Fig (3a):  The photon spheres locations at $ (r, \theta)=(2.285600868252, 1.57)$ and $ (r, \theta)=(2.305792105113,1.57)$  with respect to $( \Xi = 0.2488, \alpha = 0.39, m = 1 )$ in the $(r-\theta)$ plane of the normal vector field $n$ , (3b): the topological potential H(r) for  NCEGB BH  }}
 \label{3}
\end{center}
\end{figure}
 In the horizonless region, with $\Xi$ = 0.2488, we can see the appearance of two photon spheres with topological charges of -1 and 1, Fig. (\ref{3a}), which indicating a spacetime with TTC=0. This situation is corresponded to the existence of the minimum and maximum in the H diagram, as we can see in the Fig. (\ref{3b}). All of the above statements mean that in this case we will be dealing with a naked singularity.
 \begin{center}
\textbf{$\alpha = 0.39$ }
\end{center}
\begin{center}
\begin{table}[H]
  \centering
\begin{tabular}{|p{3cm}|p{4cm}|p{5cm}|p{1.5cm}|p{2cm}|}
  \hline
  \centering{NCEGB BH}  & \centering{Fix parametes} &\centering{Conditions}& *TTC&\ $(R_{PLPS})$\\[3mm]
   \hline
  \centering{unstable photon sphere} & \centering $ \alpha = 0.39, m = 1  $ & \centering{$0<\Xi \leq 0.1081 $} & \centering $-1$&\ $2.562080591$\\[3mm]
   \hline
 \centering{naked singularity} & \centering $\alpha = 0.39, m = 1 $ & \centering{$0.1081<\Xi \leq 0.2488 $} &\centering $0$&\ $-$ \\[3mm]
   \hline
   \centering{*Unauthorized area} & \centering $\alpha = 0.39, m = 1 $ & \centering{$\Xi> 0.2488 $} & \centering $ nothing $ &\ $-$ \\[3mm]
   \hline
   \end{tabular}
   \caption{*Unauthorized region: The region with negative or imaginary roots of $\varphi$.\\ $R_{PLPS}$: the minimum or maximum possible radius for the appearance of an unstable photon sphere.}\label{1}
\end{table}
 \end{center}
\begin{center}
\textbf{$\alpha = 0.5$ }
\end{center}
\begin{center}
\begin{table}[H]
  \centering
\begin{tabular}{|p{3cm}|p{4cm}|p{4cm}|p{4cm}|}
  \hline
  \centering{NCEGB BH }  & \centering{Fix parametes} &\centering{Conditions}&*TTC\\[3mm]
   \hline
  \centering{naked singularity} & \centering $ \alpha = 0.5, m = 1 $ & \centering{$0< \Xi \leq 0.1974$} & $0$\\[3mm]
   \hline
  \centering{*Unauthorized area} & \centering $ \alpha = 0.5, m = 1 $ & \centering{$ 0.1974 < \Xi $} &  $nothing$\\[3mm]
   \hline
      \end{tabular}
   \caption{*Unauthorized region: The region with negative or imaginary roots of $\varphi$.}\label{2}
\end{table}
 \end{center}
The results of the allowable range for the parameter $\Xi$ are given in Tables 1, 2, and 5. As stated, we see that for $\Xi$ = 0.5 the model lacks a black hole structure and only appears in the form of a naked singularity.
\subsection{TCOs}
According to Eq. (\ref{(6)}) and Eq. (\ref{(7)}) and Eq. (\ref{(11)}) for this model we will have:
\begin{equation}\label{(19)}
\mathbb{A} =E^{2} r^{2}+\frac{L^{2} \left(r^{2} \sqrt{\frac{16 \sqrt{\pi}\, \mathrm{erf}\left(\frac{\sqrt{\frac{r^{2}}{\Xi}}}{2}\right) \alpha  m +\sqrt{\pi}\, r^{3}-16 {\mathrm e}^{-\frac{r^{2}}{4 \Xi}} \sqrt{\frac{r^{2}}{\Xi}}\, \alpha  m}{r^{3}}}-\left(r^{2}+4 \alpha \right) \pi^{\frac{1}{4}}\right)}{4 \pi^{\frac{1}{4}} \alpha}.
\end{equation}
\begin{equation}\label{(20)}
\mathbb{B} =-\frac{r^{2} \left(r^{2} \sqrt{\frac{16 \sqrt{\pi}\, \mathrm{erf}\left(\frac{\sqrt{\frac{r^{2}}{\Xi}}}{2}\right) \alpha  m +\sqrt{\pi}\, r^{3}-16 {\mathrm e}^{-\frac{r^{2}}{4 \Xi}} \sqrt{\frac{r^{2}}{\Xi}}\, \alpha  m}{r^{3}}}-\left(r^{2}+4 \alpha \right) \pi^{\frac{1}{4}}\right)}{4 \pi^{\frac{1}{4}} \alpha}.
\end{equation}
\begin{equation*}\label{(0)}
b_{1}=\pi^{\frac{1}{4}} \sqrt{\frac{16 \sqrt{\pi}\, \mathrm{erf}\! \left(\frac{\sqrt{\frac{r^{2}}{\Xi}}}{2}\right) \alpha  m +\sqrt{\pi}\, r^{3}-16 {\mathrm e}^{-\frac{r^{2}}{4 \Xi}} \sqrt{\frac{r^{2}}{\Xi}}\, \alpha  m}{r^{3}}}\, \Xi  r ,
\end{equation*}
\begin{equation}\label{(21)}
\beta =\frac{-3 \mathrm{erf}\! \left(\frac{\sqrt{\frac{r^{2}}{\Xi}}}{2}\right) \sqrt{\pi}\, \Xi  m +\frac{\sqrt{\frac{r^{2}}{\Xi}}\, m \left(r^{2}+6 \Xi \right) {\mathrm e}^{-\frac{r^{2}}{4 \Xi}}}{2}+b_{1}}{\pi^{\frac{1}{4}} \sqrt{\frac{16 \sqrt{\pi}\, \mathrm{erf}\left(\frac{\sqrt{\frac{r^{2}}{\Xi}}}{2}\right) \alpha  m +\sqrt{\pi}\, r^{3}-16 {\mathrm e}^{-\frac{r^{2}}{4 \Xi}} \sqrt{\frac{r^{2}}{\Xi}}\, \alpha  m}{r^{3}}}\, \Xi  r},
\end{equation}
where erf is The Error Function. In the black hole forme we have,Fig. (\ref{4})
\begin{figure}[H]
 \begin{center}
 \subfigure[]{
 \includegraphics[height=5.5cm,width=6cm]{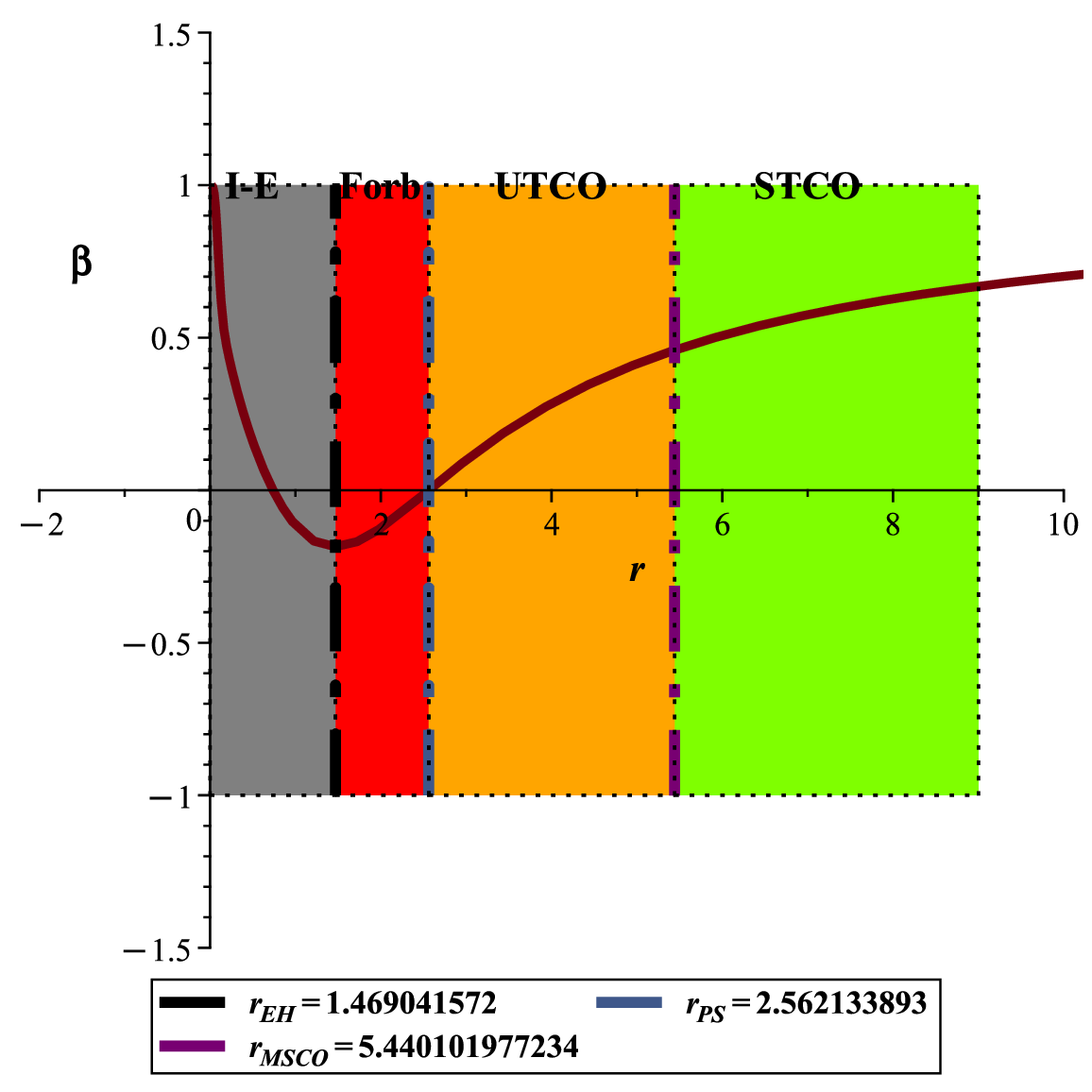}
 \label{4a}}
 \subfigure[]{
 \includegraphics[height=5.5cm,width=6cm]{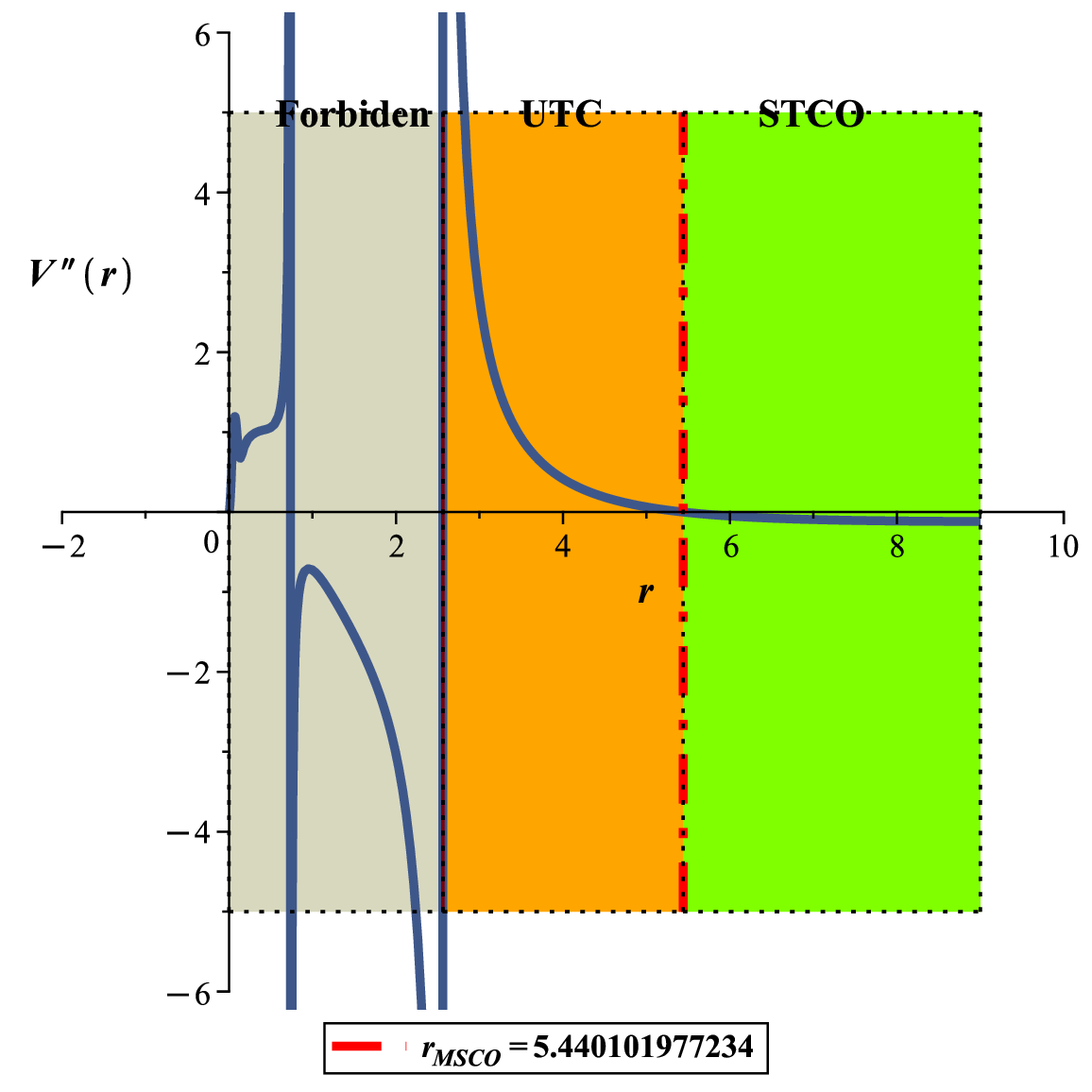}
 \label{4b}}
   \caption{\small{ With respect to $ \Xi = 0.001, \alpha = 0.39, m = 1$, Fig (4a): the $\beta$ diagram in the black hole form and (4b): The space classification  and MSCO location in the black hole mode}}
 \label{4}
\end{center}
\end{figure}
As illustrated in Fig. (\ref{4a}) and Fig. (\ref{4b}), the model in the form of a black hole continues to follow the pattern proposed in \cite{22}. Specifically, the region between the event horizon and the unstable photon sphere remains a forbidden zone for the emergence of TCOs due to the negative beta \cite{1}. Beyond the photon sphere, we should sequentially observe the regions of STCOs and UTCOs.\\ In the form of naked singularity for the TCO's we have,Fig. (\ref{5})
 \begin{figure}[H]
 \begin{center}
 \subfigure[]{
 \includegraphics[height=5cm,width=5cm]{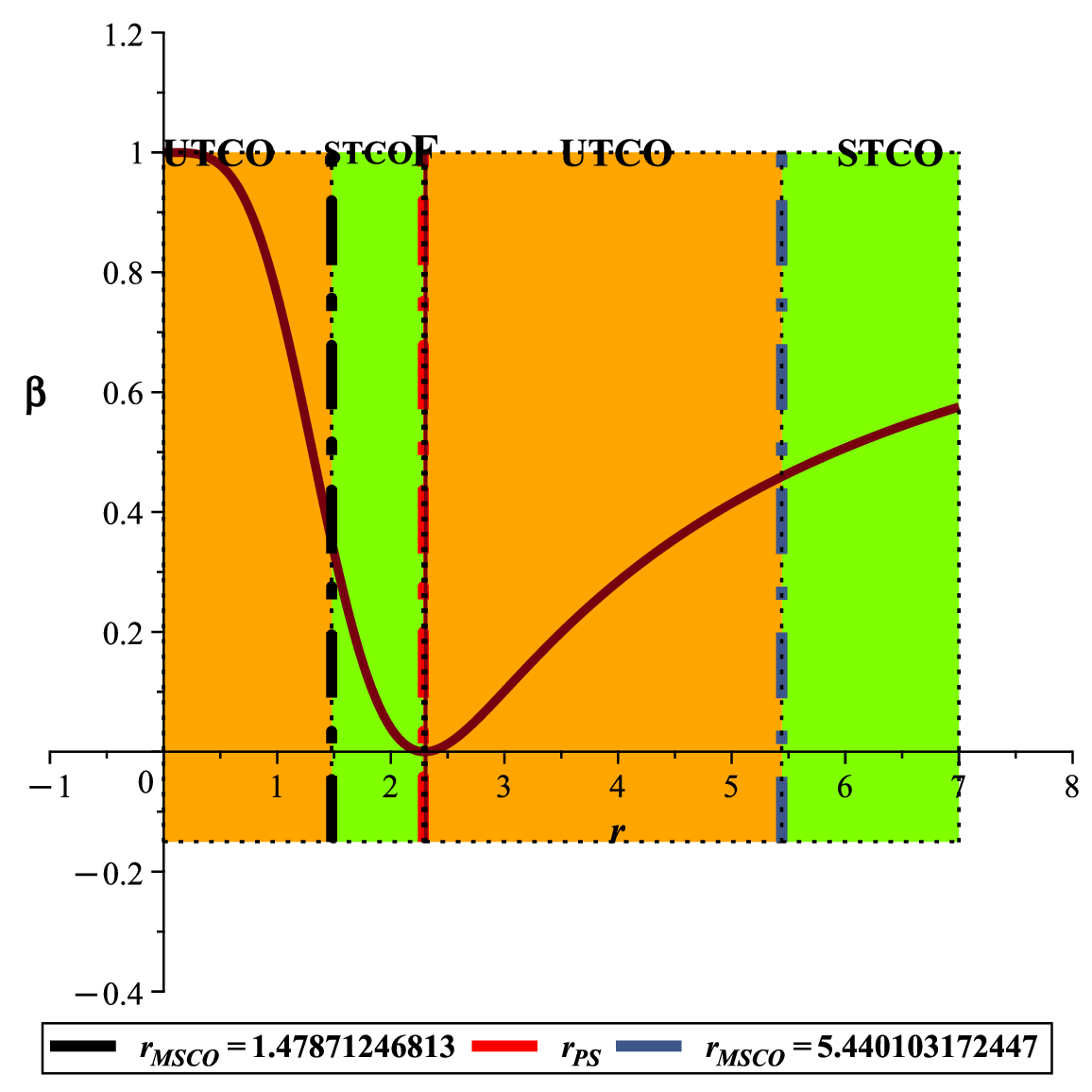}
 \label{5a}}
 \subfigure[]{
 \includegraphics[height=5cm,width=5cm]{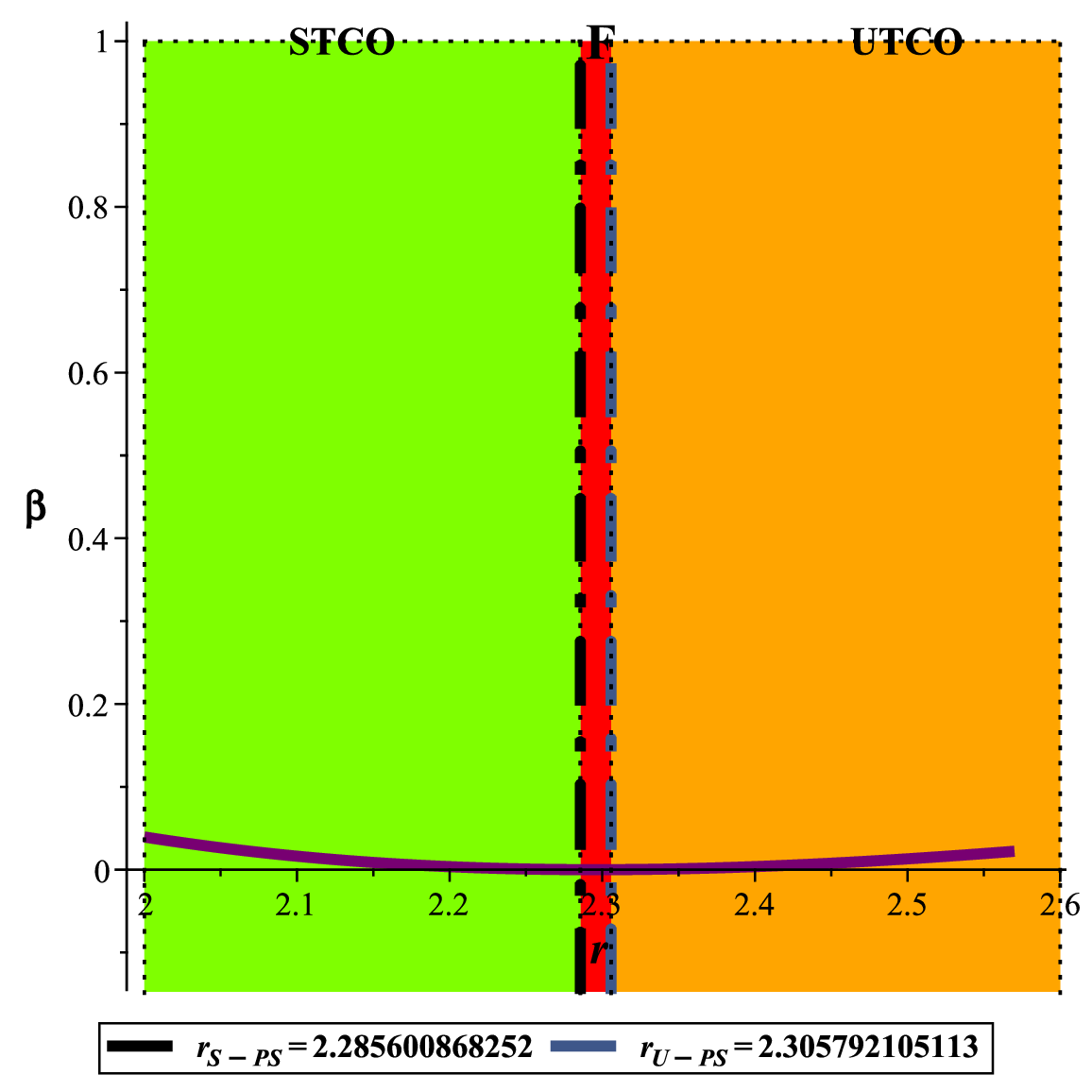}
 \label{5b}}
\subfigure[]{
 \includegraphics[height=5cm,width=5cm]{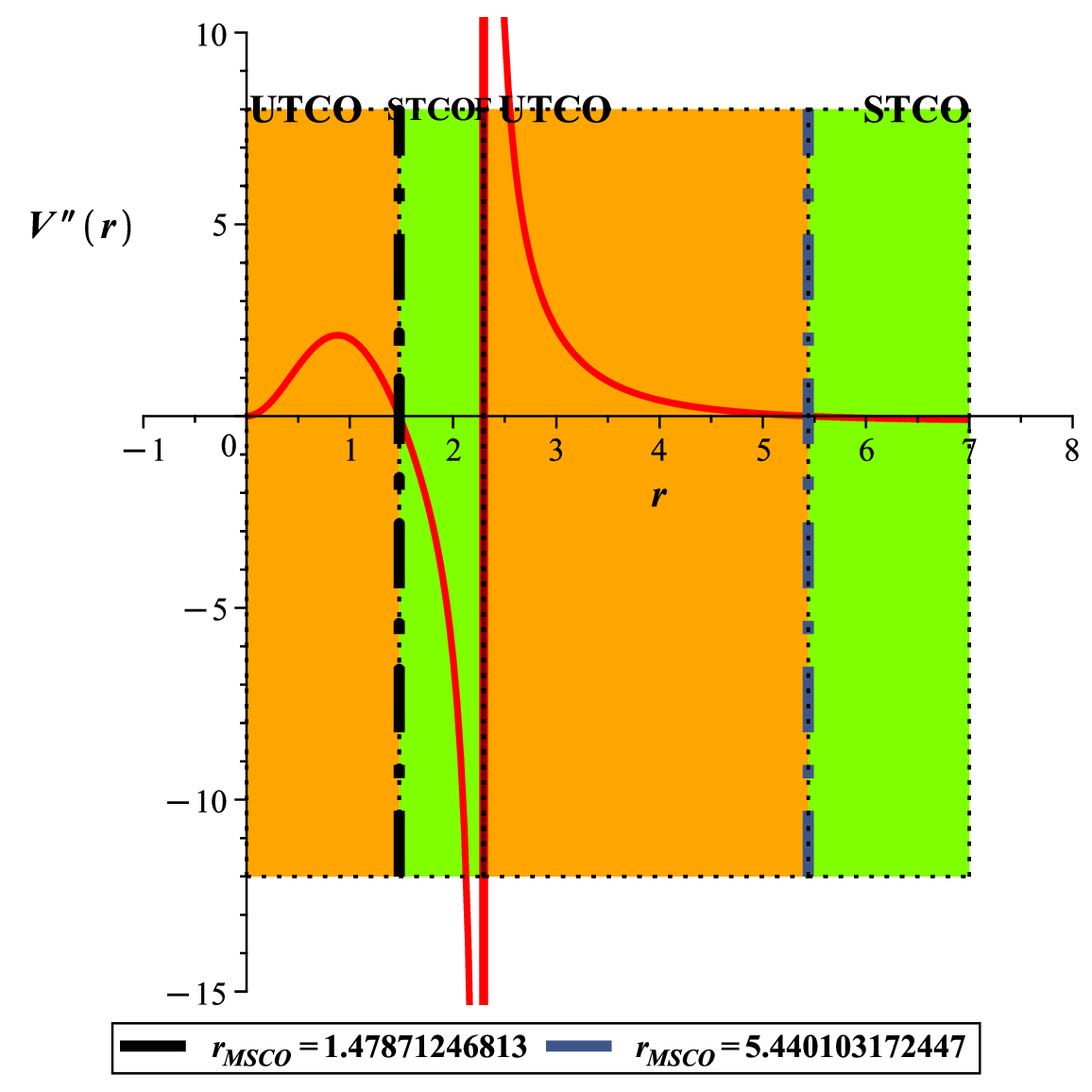}
 \label{5c}}
   \caption{\small{ With respect to $ \Xi = 0.2488, \alpha = 0.39, m = 1$, Fig (5a): the $\beta$ diagram for structure in naked singularity form , (5b): Enlarging on the forbidden area in the diagram $\beta$ and (5c): The space classification and MSCO location in the naked singularity mode}}
 \label{5}
\end{center}
\end{figure}
As we can see in Fig. (\ref{5a}) and Fig. (\ref{5c}), the model in the event horizon-free state follows the ultra-compact gravitational object model proposed in \cite{22}, that is, the boundary between the stable and unstable photospheres is the disallowed region for the emergence of TCOs due to negative beta, and after the photospheres, we should witness the boundaries of STCOs and UTCOs, respectively.
\section{Non-Commutative CHARGED Gauss-Bonnet Black Hole with CLOUD OF STRINGS }
\begin{figure}[H]
 \begin{center}
 \subfigure[]{
 \includegraphics[height=5.5cm,width=6cm]{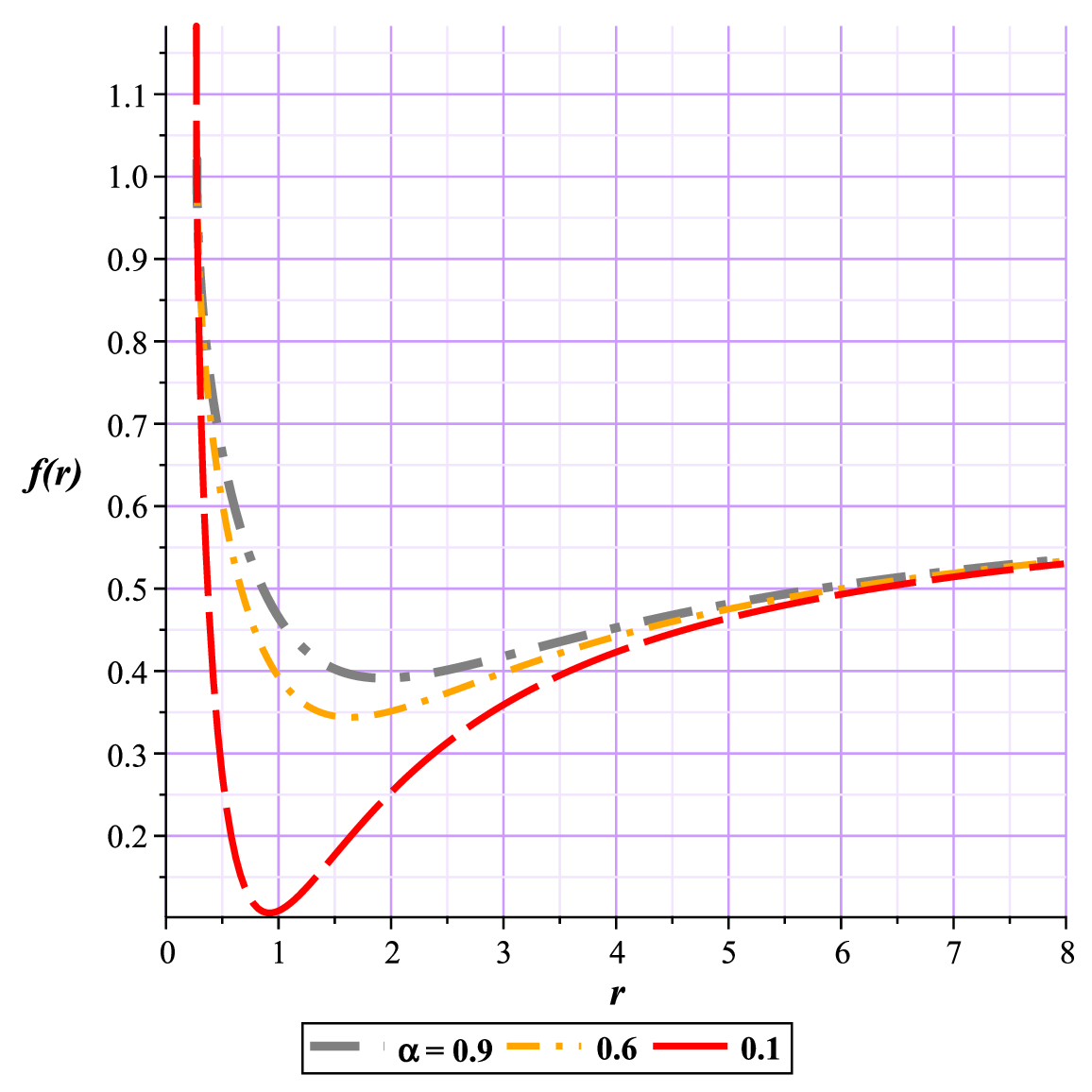}
 \label{6a}}
 \subfigure[]{
 \includegraphics[height=5.5cm,width=6cm]{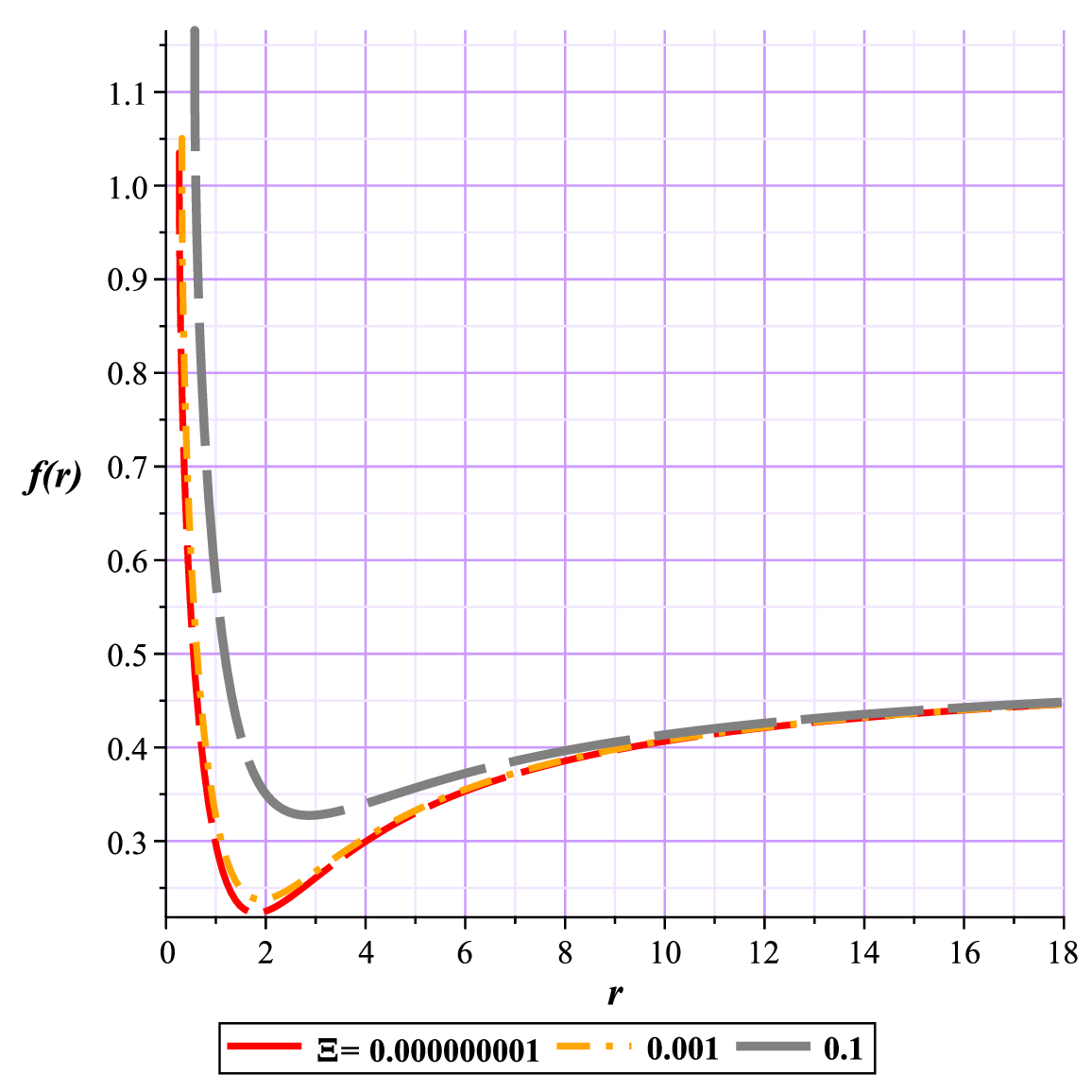}
 \label{6b}}
  \caption{\small{Fig (6a): Metric function with $ a=1, \Xi = 10^{-7}$ and different $\alpha$ ,  Fig (6b) $a=1,\alpha=0.5$,  and different $\Xi$ for Charged NC model with Cloud of Strings. }}
 \label{6}
\end{center}
\end{figure}

In string theory, elementary particles are represented as one-dimensional strings that vibrate at different frequencies. The “Strings Cloud” concept represents a scenario where a black hole is surrounded or enveloped by a collection of strings. This cloud can be seen as a statistical ensemble of many strings interacting with the black hole.
The metric for such a black hole is \cite{26}
\begin{equation}\label{(22)}
f \! \left(r \right)=1+\frac{\left(1-\sqrt{1+4 \left(\frac{4 \left(\arctan \left(\frac{r}{\sqrt{\Xi  \pi}}\right)-\frac{\sqrt{\Xi  \pi}\, r}{\Xi  \pi +r^{2}}\right) m}{\pi  r^{3}}-\frac{q}{r^{4}}+\frac{a}{r^{2}}\right) \alpha}\right) r^{2}}{4 \alpha}
\end{equation}
where $\alpha$ is the GB coupling constant, m is the mass, q is the charge, and a is the cloud of string parameter, which is considered positive. Our studies indicate that by adding charge and the string cloud parameter, the constraints on $\alpha$ seem to no longer exist, and at first glance, it appears that the value of $\alpha$ can change freely. However, since this model is influenced by multiple parameters, the variations of other parameters must be examined more carefully. For instance, the string cloud parameter is typically considered to be between 0 and 1 in most articles. Nevertheless, it can be easily seen here that to maintain a black hole in sub-extremal conditions $Q < M$ and given the above limit for the string cloud parameter ($a < 1$), the role of this quantity in determining the permissible range for other parameters becomes significantly prominent, So that only those ranges that structurally differ from each other will maintain the model in the form of a black hole.
For instance, with m = 1 and the Gauss-Bonnet parameter $\alpha > 0.1$, the structure will practically be in the form of a naked singularity in all regions (Fig. (\ref{6})).
Therefore, given that the main focus of this study is on the interaction between the NC and Gauss-Bonnet parameters, our research will be based on identifying these different ranges concerning the Gauss-Bonnet parameter and we examined the parameter $\alpha$ in three distinct regions.
\subsection{$ Case I:\alpha < 0 $ }
Since this choice is uncommon, the first question that arises is whether a negative $\alpha$ has any physical meaning. To address this, it should be noted that when we refer to negative Gauss-Bonnet parameters, we are discussing scenarios where these curvature terms have negative coefficients. This can lead to intriguing effects in the context of gravity and cosmology \cite{27,28}. For instance, in some 4D models, a negative Gauss-Bonnet parameter is essential to ensure the stability and consistency of the solutions. It can help avoid certain instabilities that might otherwise arise in the gravitational field equations \cite{27}. Additionally, studies on the absorption of particles by black holes in Gauss-Bonnet gravity have indicated that under specific conditions, the parameter might turn negative when considering the dynamics of massless and charged particles \cite{29}. Also, in modified theories of gravity, particularly in 4D Einstein-Gauss-Bonnet gravity, the parameter can become negative through the rescaling of the coupling parameters involved. This rescaling is necessary to construct self-consistent theoretical frameworks \cite{27}. Perhaps the most notable and fascinating example is the mimicry of dark energy. Negative Gauss-Bonnet parameters can influence the dynamics of the universe in a way that resembles the effects of dark energy. Essentially, they can produce a similar accelerated expansion without the need to invoke a separate dark energy component. This implies that the observed acceleration of the universe could be explained by modifications to gravity itself rather than an unknown energy form \cite{30}.
\begin{center}
\textbf{Topological Photon Sphere }
\end{center}
With respect to Eq. (\ref{(4)}), Eq. (\ref{(5)}),Eq. (\ref{(17)}),Eq. (\ref{(18)}) we have:
\begin{equation}\label{(23)}
\begin{split}
\mathbb{H} =\frac{\sqrt{4+\frac{\left(1-\sqrt{1+4 \left(\frac{4 \left(\arctan \left(\frac{r}{\sqrt{\Xi  \pi}}\right)-\frac{\sqrt{\Xi  \pi}\, r}{\Xi  \pi +r^{2}}\right) m}{\pi  r^{3}}-\frac{q}{r^{4}}+\frac{a}{r^{2}}\right) \alpha}\right) r^{2}}{\alpha}}}{2 \sin \! \left(\theta \right) r}.
\end{split}
\end{equation}
\begin{equation*}\label{(0)}
\begin{split}
\varphi_{0}=\sqrt{\frac{16 m r \alpha  \left(\Xi  \pi +r^{2}\right) \arctan \! \left(\frac{r}{\sqrt{\pi}\, \sqrt{\Xi}}\right)-16 \sqrt{\pi}\, \sqrt{\Xi}\, m \alpha  r^{2}+4 \left(\Xi  \pi +r^{2}\right) \pi  \left(a \alpha  r^{2}+\frac{1}{4} r^{4}-\alpha  q \right)}{r^{4} \left(\Xi  \pi +r^{2}\right)}},
\end{split}
\end{equation*}
\begin{equation*}\label{(0)}
\begin{split}
\varphi_{1}=\varphi_{0} \left(-2 \pi^{3} \Xi^{\frac{5}{2}} r^{2}-4 \pi^{2} \Xi^{\frac{3}{2}} r^{4}-2 \pi  \sqrt{\Xi}\, r^{6}\right),
\end{split}
\end{equation*}
\begin{equation*}\label{(0)}
\begin{split}
\varphi_{2}=6 m r \left(\sqrt{\pi}\, \sqrt{\Xi}\, r^{4}+2 \pi^{\frac{3}{2}} r^{2} \Xi^{\frac{3}{2}}+\pi^{\frac{5}{2}} \Xi^{\frac{5}{2}}\right) \arctan \! \left(\frac{r}{\sqrt{\pi}\, \sqrt{\Xi}}\right)+2 r^{2} \pi^{\frac{5}{2}} \left(a \,r^{2}-2 q \right) \Xi^{\frac{3}{2}}+\pi^{\frac{7}{2}} \left(a \,r^{2}-2 q \right) \Xi^{\frac{5}{2}},
\end{split}
\end{equation*}
\begin{equation*}\label{(0)}
\begin{split}
\varphi_{3}=\sqrt{\frac{16 m r \alpha  \left(\Xi  \pi +r^{2}\right) \arctan \! \left(\frac{r}{\sqrt{\pi}\, \sqrt{\Xi}}\right)-16 \sqrt{\pi}\, \sqrt{\Xi}\, m \alpha  r^{2}+4 \left(\Xi  \pi +r^{2}\right) \pi  \left(a \alpha  r^{2}+\frac{1}{4} r^{4}-\alpha  q \right)}{r^{4} \left(\Xi  \pi +r^{2}\right)}},
\end{split}
\end{equation*}
\begin{equation}\label{(24)}
\begin{split}
\varphi_{r}=\frac{\csc \! \left(\theta \right) \left(\varphi_{1}+\varphi_{2}+\left(\sqrt{\Xi}\, r^{2} \left(a \,r^{2}-2 q \right) \pi^{\frac{3}{2}}-10 m \,r^{2} \pi  \Xi -6 m \pi^{2} \Xi^{2}\right) r^{2}\right)}{2 \sqrt{\Xi}\, \varphi_{3} \pi  \left(\Xi  \pi +r^{2}\right)^{2} r^{4}}.
\end{split}
\end{equation}
\begin{equation}\label{(25)}
\begin{split}
\varphi_{\theta}=-\frac{\sqrt{4+\frac{\left(1-\sqrt{1+4 \left(\frac{4 \left(\arctan \left(\frac{r}{\sqrt{\Xi  \pi}}\right)-\frac{\sqrt{\Xi  \pi}\, r}{\Xi  \pi +r^{2}}\right) m}{\pi  r^{3}}-\frac{q}{r^{4}}+\frac{a}{r^{2}}\right) \alpha}\right) r^{2}}{\alpha}}\, \cos \! \left(\theta \right)}{2 \sin \! \left(\theta \right)^{2} r^{2}}.\\\\
\end{split}
\end{equation}
\begin{figure}[H]
 \begin{center}
 \subfigure[]{
 \includegraphics[height=5cm,width=5cm]{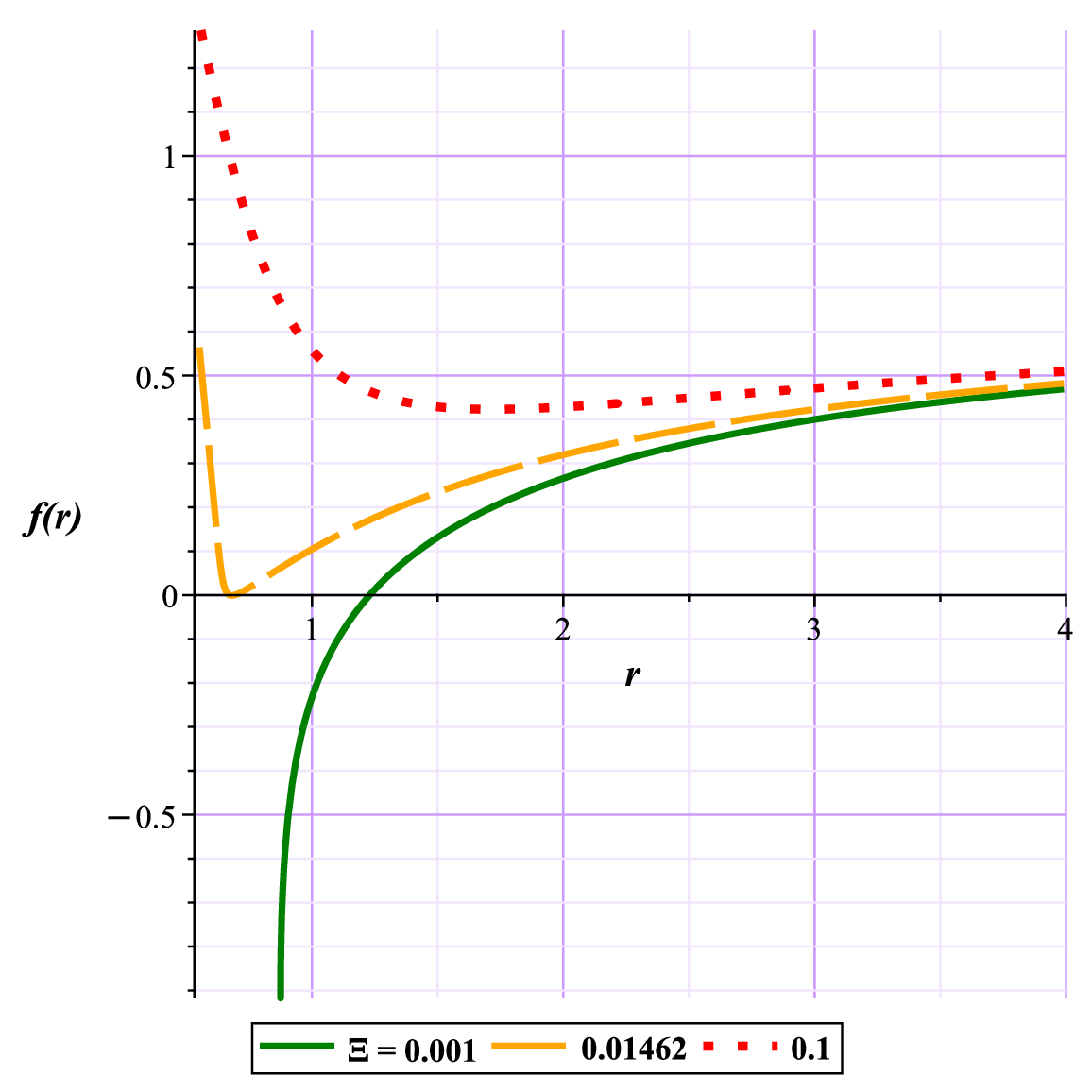}
 \label{7a}}
 \subfigure[]{
 \includegraphics[height=5cm,width=5cm]{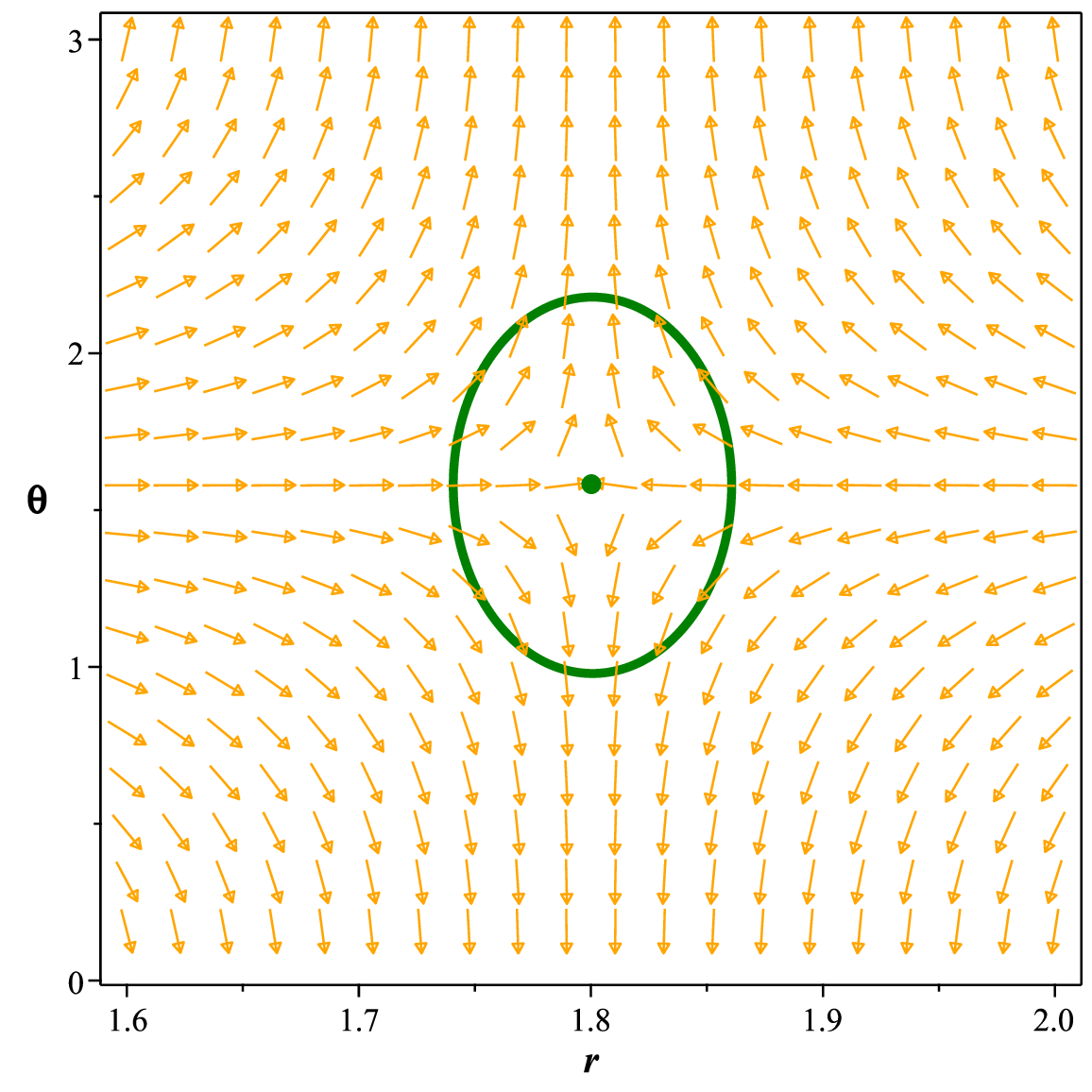}
 \label{7b}}
\subfigure[]{
 \includegraphics[height=5cm,width=5cm]{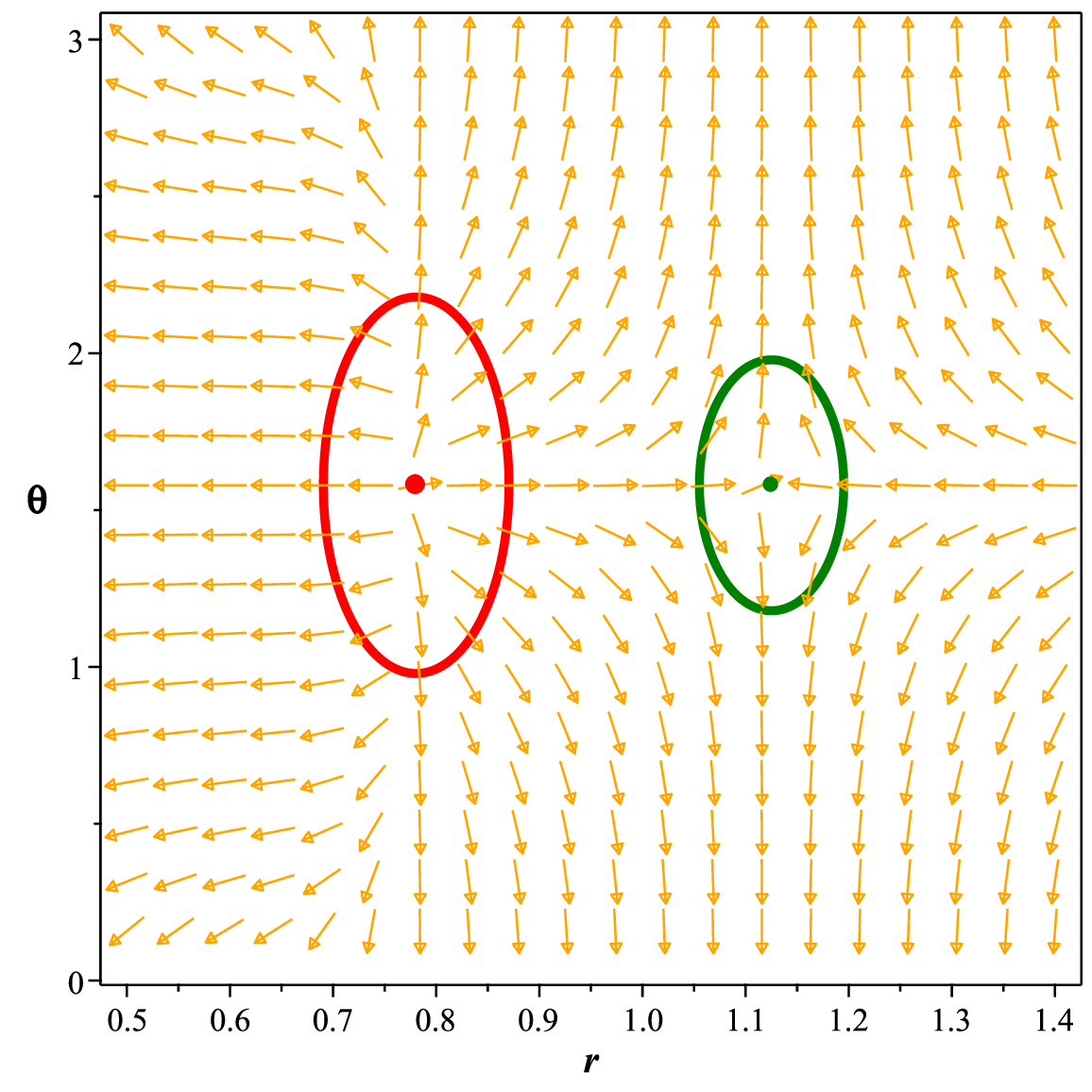}
 \label{7c}}
   \caption{\small{Fig (7a): Metric function with $a=0.6,\alpha=-0.1$  and different $\Xi$ for Black Hole with Cloud of strings, Fig (7b): The photon sphere is location at $ (r, \theta)=(1.800667715288,1.57)$ with respect to $(a = 0.6, q = 0.6, \Xi = 0.001, \alpha = -0.1, m = 1 )$, (7c): The photon spheres location at $ (r,\theta)=(0.78056312953, 1.57)$ and $ (r, \theta)=(1.125202804843, 1.57)$ with respect to $(a = 1.4, q = 0.6, \Xi = 0.0155, \alpha = -0.1, m = 1 )$ for naked singularity with CLOUD OF STRINGS}}
 \label{7}
\end{center}
\end{figure}
In Fig. (\ref{7a}), we observe that there is always a critical $\Xi$ beyond which the model takes the form of a naked singularity. For the selected values, an extremal black hole appears at $\Xi = 0.01462$. In Figures Fig. (\ref{7b}) and Fig. (\ref{7c}), we can see TTC = -1 and TTC = 0, respectively.
\begin{center}
\begin{table}[H]
  \centering
\begin{tabular}{|p{3cm}|p{4cm}|p{5cm}|p{1.5cm}|p{2cm}|}
  \hline
  \centering{NC-CLOUD OF STRINGS BH}  & \centering{Fix parametes} &\centering{Conditions}& *TTC&\ $(R_{PLPS})$\\[3mm]
   \hline
  \centering{unstable photon sphere} & \centering $ a = 0.6, q = 0.6,\alpha = -0.1, m = 1  $ & \centering{$0<\Xi \leq 0.01462 $} & \centering $-1$&\ $1.1925029$\\[3mm]
   \hline
 \centering{naked singularity} & \centering $ a = 0.6, q = 0.6, \alpha = -0.1, m = 1 $ & \centering{$0.01462<\Xi \leq 0.01641 $} &\centering $0$&\ $-$ \\[3mm]
   \hline
   \centering{*Unauthorized area} & \centering $ a = 0.6, q = 0.6, \alpha = -0.1, m = 1 $ & \centering{$\Xi> 0.01641 $} & \centering $ nothing $ &\ $-$ \\[3mm]
   \hline
   \end{tabular}
   \caption{*Unauthorized region: The region with negative or imaginary roots of $\varphi$.\\ $R_{PLPS}$: the minimum or maximum possible radius for the appearance of an unstable photon sphere.}\label{3}
\end{table}
 \end{center}
 The results are presented in full in Table (3).
\begin{center}
\textbf{TCOs }
\end{center}
According to Eq. (\ref{(6)}) and Eq. (\ref{(7)}) and Eq. (\ref{(11)}) for this model we will have:
\begin{equation}\label{(26)}
\mathbb{A} =E^{2} r^{2}-L^{2} \left(1+\frac{\left(1-\sqrt{1+4 \left(\frac{4 \left(\arctan \left(\frac{r}{\sqrt{\Xi  \pi}}\right)-\frac{\sqrt{\Xi  \pi}\, r}{\Xi  \pi +r^{2}}\right) m}{\pi  r^{3}}-\frac{q}{r^{4}}+\frac{a}{r^{2}}\right) \alpha}\right) r^{2}}{4 \alpha}\right).
\end{equation}
\begin{equation}\label{(27)}
\mathbb{B} =r^{2} \left(1+\frac{\left(1-\sqrt{1+4 \left(\frac{4 \left(\arctan \left(\frac{r}{\sqrt{\Xi  \pi}}\right)-\frac{\sqrt{\Xi  \pi}\, r}{\Xi  \pi +r^{2}}\right) m}{\pi  r^{3}}-\frac{q}{r^{4}}+\frac{a}{r^{2}}\right) \alpha}\right) r^{2}}{4 \alpha}\right).
\end{equation}
\begin{equation*}\label{(0)}
\begin{split}
&b_{1}=-6 r \left(\frac{r^{4} \sqrt{\pi}}{2}+\pi^{\frac{3}{2}} r^{2} \Xi +\frac{\pi^{\frac{5}{2}} \Xi^{2}}{2}\right) m \arctan \! \left(\frac{r}{\sqrt{\pi}\, \sqrt{\Xi}}\right)+\left(-\frac{1}{2} a \,r^{6}+q \,r^{4}\right) \pi^{\frac{3}{2}}\\
&-\Xi  r^{2} \left(a \,r^{2}-2 q \right) \pi^{\frac{5}{2}}-\frac{\Xi^{2} \left(a \,r^{2}-2 q \right) \pi^{\frac{7}{2}}}{2},\\
\end{split}
\end{equation*}
\begin{equation*}\label{(0)}
\begin{split}
&b_{2}=\sqrt{\frac{16 m r \alpha  \left(\Xi  \pi +r^{2}\right) \arctan \! \left(\frac{r}{\sqrt{\pi}\, \sqrt{\Xi}}\right)-16 \sqrt{\pi}\, \sqrt{\Xi}\, m \alpha  r^{2}+4 \pi  \left(\Xi  \pi +r^{2}\right) \left(a \alpha  r^{2}+\frac{1}{4} r^{4}-q \alpha \right)}{r^{4} \left(\Xi  \pi +r^{2}\right)}}\, \left(\Xi  \pi +r^{2}\right)^{2}+\\
&3 \Xi^{\frac{3}{2}} m \pi +5 \sqrt{\Xi}\, m \,r^{2},\\
\end{split}
\end{equation*}
\begin{equation*}\label{(0)}
b_{3}=\sqrt{\frac{16 m r \alpha  \left(\Xi  \pi +r^{2}\right) \arctan \! \left(\frac{r}{\sqrt{\pi}\, \sqrt{\Xi}}\right)-16 \sqrt{\pi}\, \sqrt{\Xi}\, m \alpha  r^{2}+4 \pi  \left(\Xi  \pi +r^{2}\right) \left(a \alpha  r^{2}+\frac{1}{4} r^{4}-q \alpha \right)}{r^{4} \left(\Xi  \pi +r^{2}\right)}}\, \pi  \left(\Xi  \pi +r^{2}\right)^{2} r^{2} ,
\end{equation*}
\begin{equation}\label{(28)}
\beta =\frac{b_{2} \pi  r^{2}+b_{1}}{b_{3}}.
\end{equation}
\begin{figure}[H]
 \begin{center}
 \subfigure[]{
 \includegraphics[height=5.5cm,width=6cm]{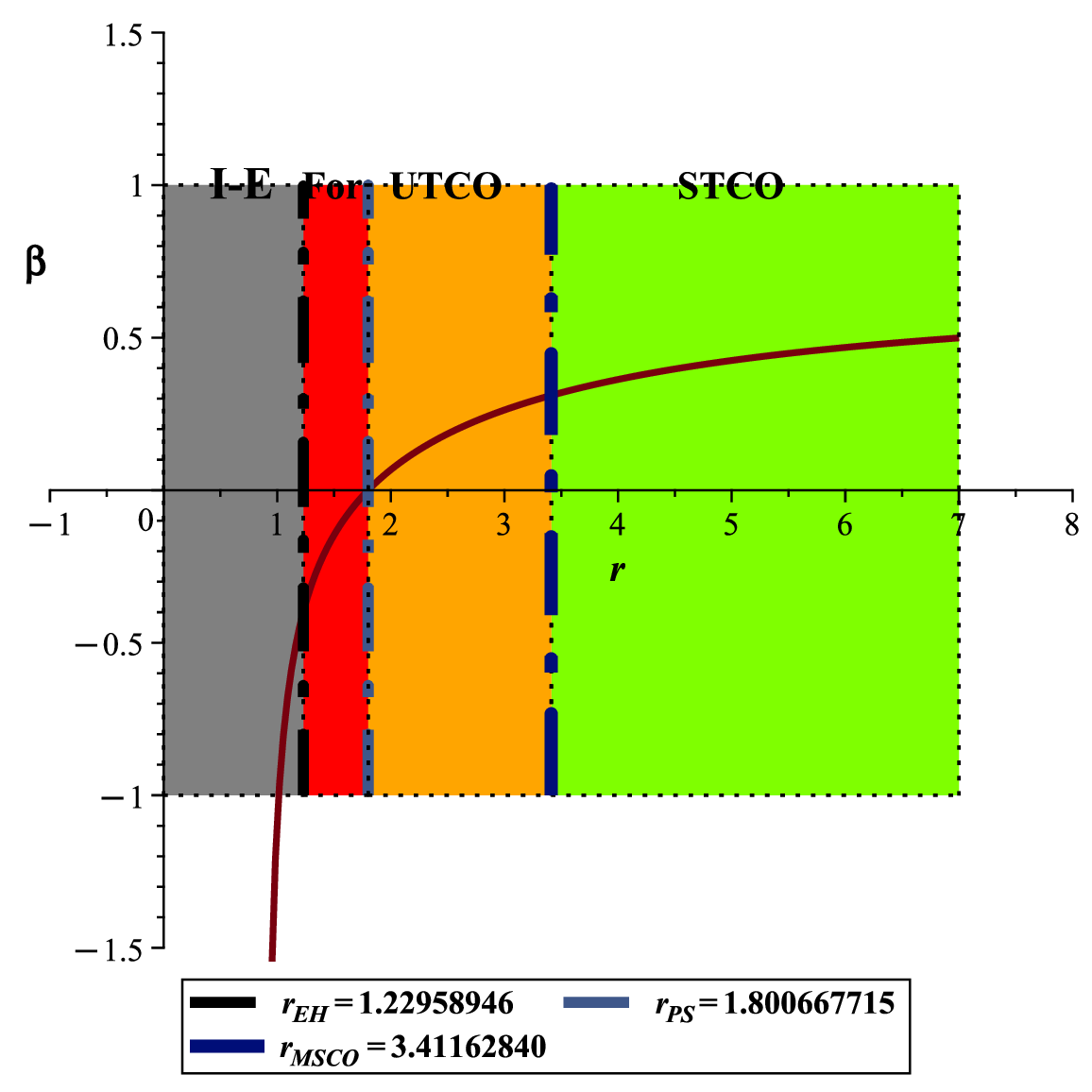}
 \label{8a}}
 \subfigure[]{
 \includegraphics[height=5.5cm,width=6cm]{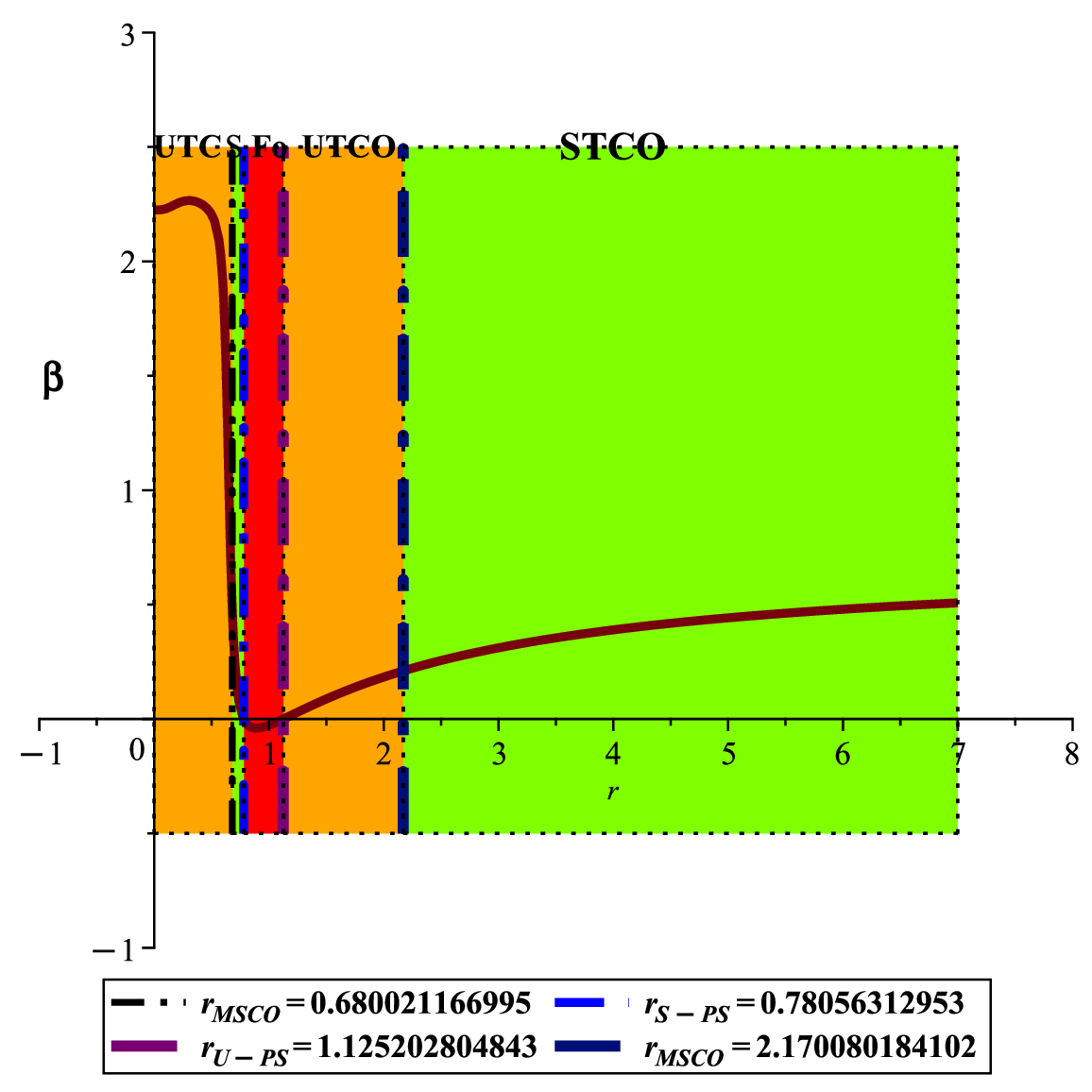}
 \label{8b}}
 \caption{\small{Fig (8a): With respect to $ a = 0.6, q = 0.6, \Xi = 0.001, \alpha = -0.1, m = 1$,  the $\beta$ diagram in the black hole form and (8b): With respect to $ a = 0.6, q = 0.6, \Xi = 0.0155, \alpha = -0.1, m = 1$, the $\beta$ diagram in the naked singularity mode}}
 \label{8}
\end{center}
\end{figure}
In this case, Fig. (\ref{8}), the structure still follows the Standard Model model \cite{22} in the black hole and naked singularity forms.
\subsection{$ Case II: 0<\alpha<0.1  $ }
One of the regions that can be distinguished is this area. In this region, while maintaining the permissible range for the string cloud parameter, alpha can be positive. However, to have a black hole, we must effectively reduce the influence of the Gauss-Bonnet term by choosing very small values for $\alpha$, below 0.1, which might not be desirable given its significance. Nevertheless, in this region, black hole behavior can be observed.
\begin{center}
\textbf{Topological Photon Sphere }
\end{center}
To avoid repetition, we will suffice with the results table.
\begin{center}
\begin{table}[H]
  \centering
\begin{tabular}{|p{3cm}|p{4cm}|p{5cm}|p{1.5cm}|p{2cm}|}
  \hline
  \centering{NC-CLOUD OF STRINGS BH}  & \centering{Fix parametes} &\centering{Conditions}& *TTC&\ $(R_{PLPS})$\\[3mm]
   \hline
  \centering{unstable photon sphere} & \centering $ a = 0.8, q = 0.6,  \alpha = 0.01, m = 1  $ & \centering{$0<\Xi \leq 0.00186 $} & \centering $-1$&\ $1.1925029$\\[3mm]
   \hline
 \centering{naked singularity} & \centering $ a = 0.8, q = 0.6, \alpha = 0.01, m = 1 $ & \centering{$0.00186<\Xi \leq 0.0045 $} &\centering $0$&\ $-$ \\[3mm]
   \hline
   \centering{*Unauthorized area} & \centering $ a = 0.8, q = 0.6, \alpha = 0.01, m = 1 $ & \centering{$\Xi> 0.0045 $} & \centering $ nothing $ &\ $-$ \\[3mm]
   \hline
   \end{tabular}
   \caption{*Unauthorized region: The region with negative or imaginary roots of $\varphi$\\ $R_{PLPS}$: the minimum or maximum possible radius for the appearance of an unstable photon sphere.}\label{4}
\end{table}
 \end{center}
A noteworthy point in Table (4) is the significant reduction in the permissible range of the parameter $\Xi$ which will be compared with other cases in the discussion section.
\begin{center}
\textbf{TCOs }
\end{center}
\begin{figure}[H]
 \begin{center}
 \subfigure[]{
 \includegraphics[height=5.5cm,width=6cm]{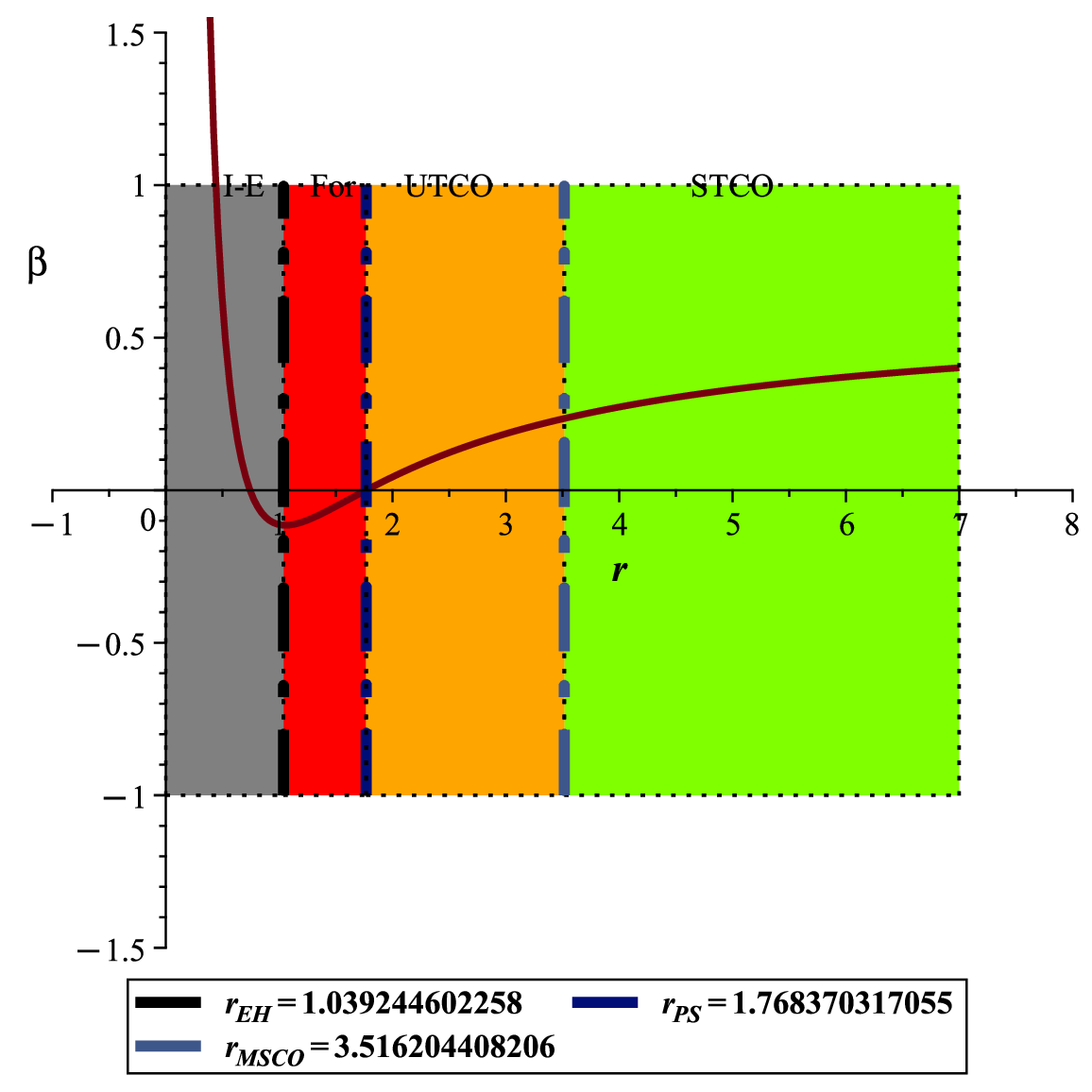}
 \label{9a}}
 \subfigure[]{
 \includegraphics[height=5.5cm,width=6cm]{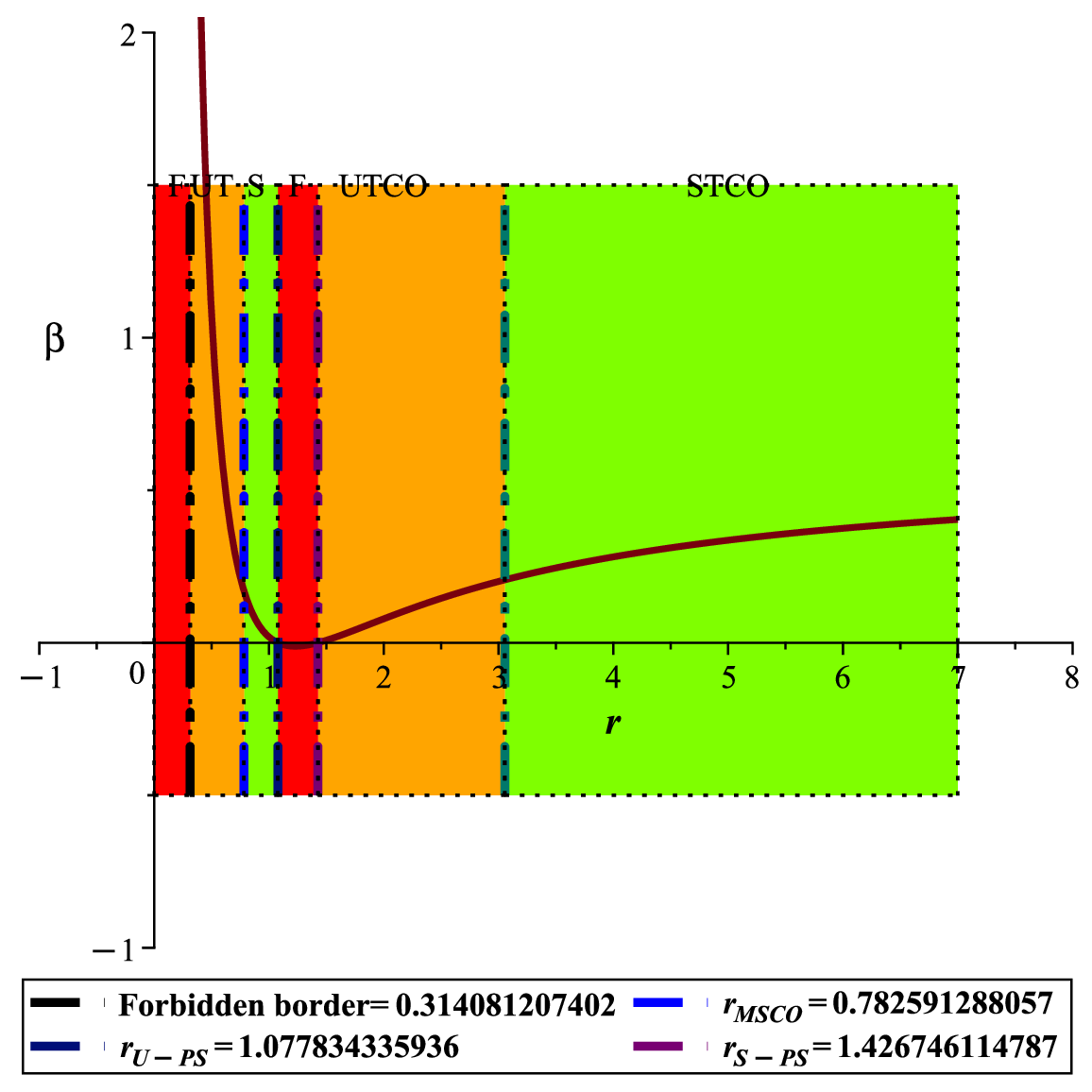}
 \label{9b}}
   \caption{\small{Fig (9a): With respect to $ a = 0.8, q = 0.6, \Xi = 0.001, \alpha = 0.01, m = 1$,  the $\beta$ diagram in the black hole form and (9b): With respect to $ a = 0.8, q = 0.6, \Xi = 0.004, \alpha = 0.01, m = 1$, the $\beta$ diagram in the naked singularity mode}}
 \label{9}
\end{center}
\end{figure}
Although in this case, the structure in the form of a black hole exhibits standard behavior (Fig. (\ref{9a}), in the naked singularity state, the addition of a forbidden zone behind the stable photon sphere and near the center is particularly interesting (Fig. (\ref{9b}). This feature has not been present in any of the models we have examined so far.
\subsection{$ Case III: \alpha> 0.1  $ }
In this region, there is a distinct difference compared to other regions, which is the sensitivity of this range to the value of  m. It appears that there is a critical mass for this range, below which the conditions necessary for forming a black hole do not exist. For example, with $ m = 1$, there is practically no region for $ \alpha < 0.1 $ where the model has an event horizon. Alternatively, for parameters such as $ a = 0.6,  q = 1 ,  \Xi = 0.001, \alpha = 0.3 $, this critical mass is $ m = 1.857 $, below which the structure takes the form of a naked singularity. This requirement for a higher mass causes this state to be radially different in magnitude from previous states, although it is still possible to make some statements about the model's accuracy based on the range of $ \Xi $. Accordingly, we chose the values $m = 3, q = 1, \alpha = 0.3, a = 0.6$
\begin{center}
\textbf{Topological Photon Sphere }
\end{center}
For these choices, the results table will be as follows:
\begin{center}
\begin{table}[H]
  \centering
\begin{tabular}{|p{3cm}|p{4cm}|p{5cm}|p{1.5cm}|p{2cm}|}
  \hline
  \centering{NC-CLOUD OF STRINGS BH}  & \centering{Fix parametes} &\centering{Conditions}& *TTC&\ $(R_{PLPS})$\\[3mm]
   \hline
  \centering{unstable photon sphere} & \centering $ a = 0.6, q = 1,  \alpha = 0.3, m = 3  $ & \centering{$0<\Xi \leq 0.1074 $} & \centering $-1$&\ $2.10419899$\\[3mm]
   \hline
 \centering{naked singularity} & \centering $ a = 0.6, q = 1,\alpha = 0.3, m = 3 $ & \centering{$0.1074<\Xi \leq 0.1602 $} &\centering $0$&\ $-$ \\[3mm]
   \hline
   \centering{*Unauthorized area} & \centering $ a = 0.6, q = 1, \alpha = 0.3, m = 3 $ & \centering{$\Xi> 0.1602 $} & \centering $ nothing $ &\ $-$ \\[3mm]
   \hline
   \end{tabular}
   \caption{*Unauthorized region: The region with negative or imaginary roots of $\varphi$.\\ $R_{PLPS}$: the minimum or maximum possible radius for the appearance of an unstable photon sphere.}\label{5}
\end{table}
 \end{center}
\begin{center}
\textbf{TCOs }
\end{center}
\begin{figure}[H]
 \begin{center}
 \subfigure[]{
 \includegraphics[height=5.5cm,width=6cm]{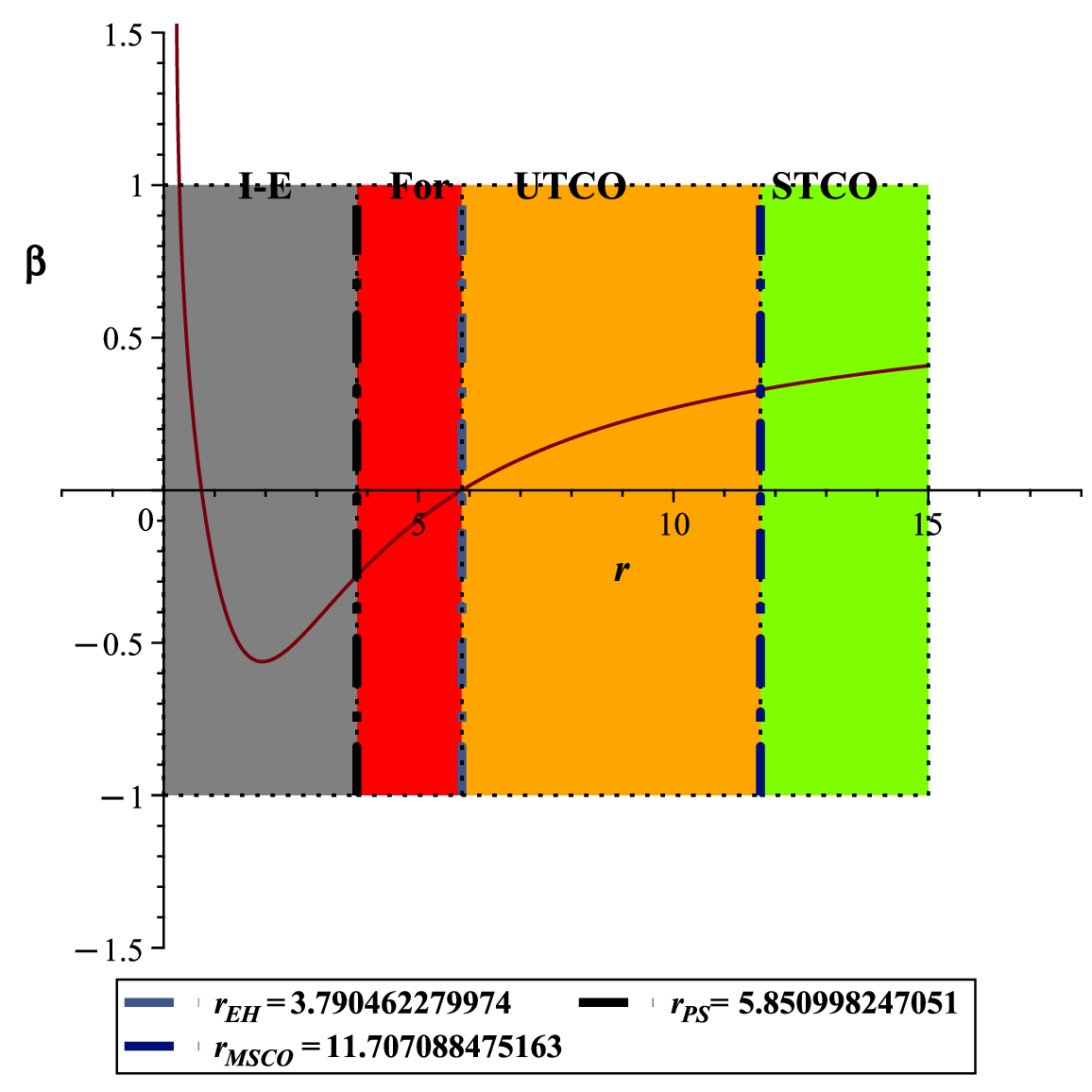}
 \label{10a}}
 \subfigure[]{
 \includegraphics[height=5.5cm,width=6cm]{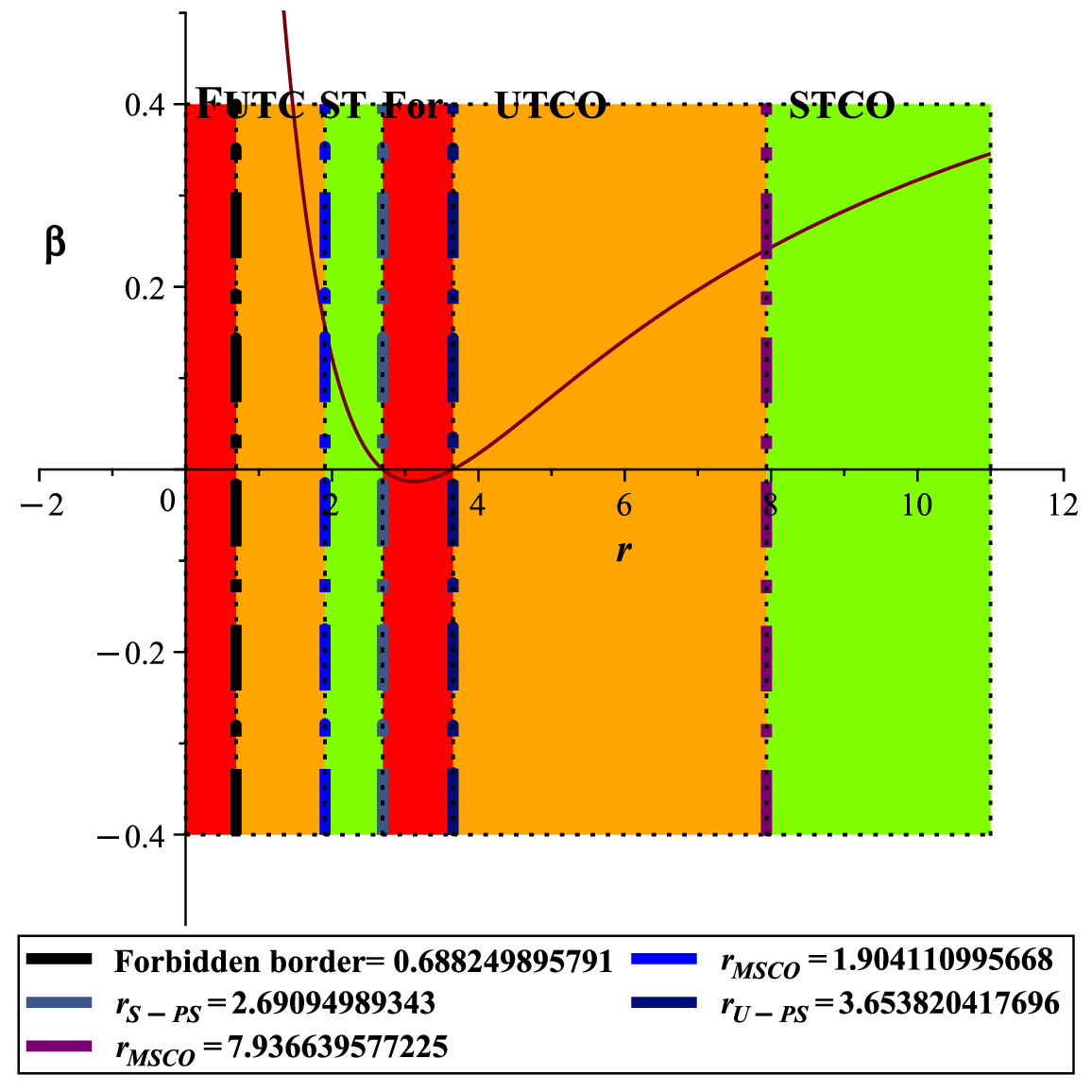}
 \label{10b}}
   \caption{\small{Fig (10a): With respect to $ a = 0.6, q = 1, \Xi = 0.001, \alpha = 0.3, m = 3$,  the $\beta$ diagram in the black hole form and (10b): With respect to $ a = 0.6, q = 1, \Xi = 0.15, \alpha = 0.3, m = 3$, the $\beta$ diagram in the naked singularity mode}}
 \label{10}
\end{center}
\end{figure}
The status of TCOs in both black hole and naked singularity states is shown in Fig. (\ref{10}). It appears that in positive alpha values, a new forbidden zone always forms near the center in the naked singularity state. Given the absence of such a zone in negative alpha values and in other models \cite{1}, the presence of this region could be an interesting difference compared to other models.
\section{SuperExtremality,Temperature and WGC as WCCC protector }
\begin{figure}[H]
 \begin{center}
 \subfigure[]{
 \includegraphics[height=5cm,width=5cm]{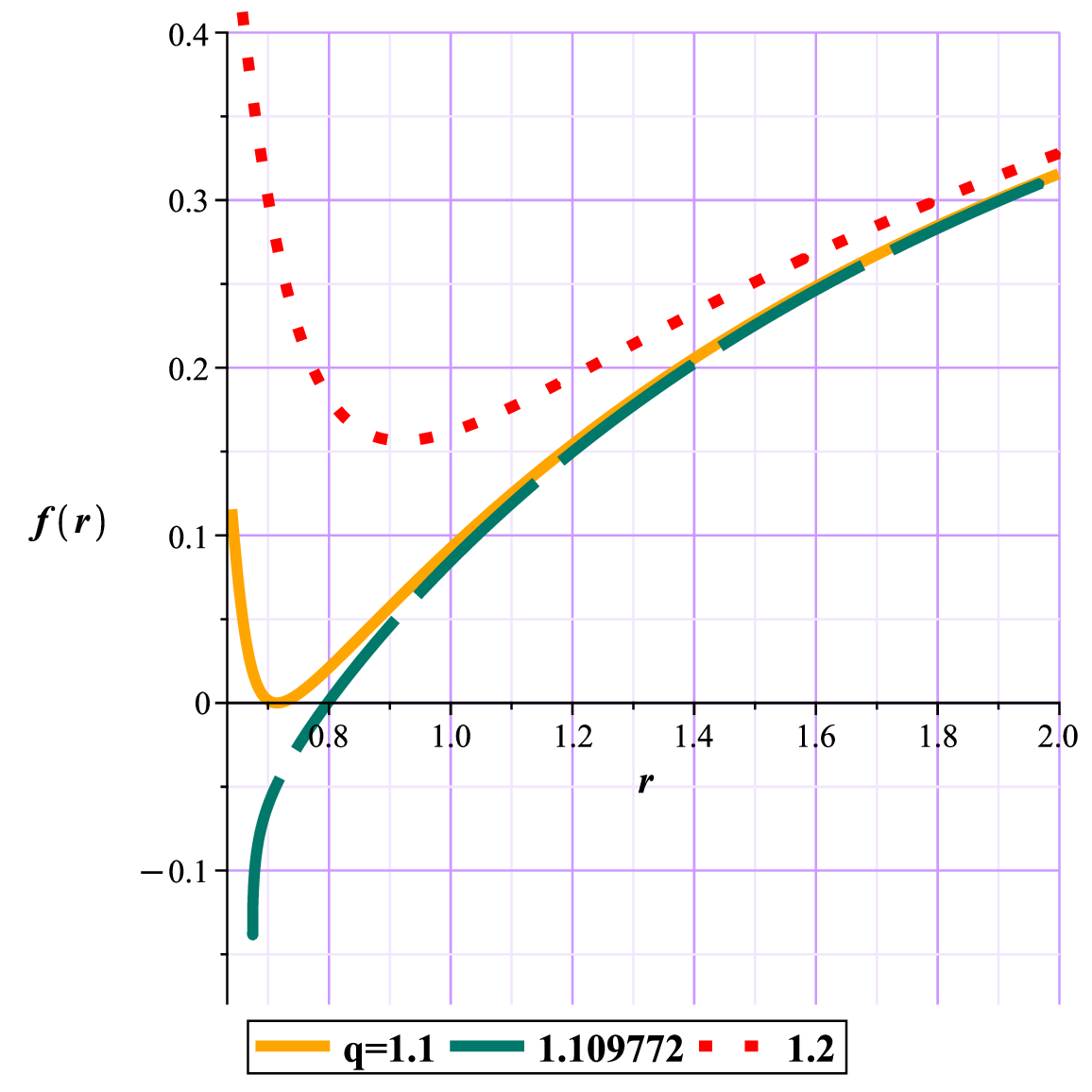}
 \label{11a}}
 \subfigure[]{
 \includegraphics[height=5cm,width=5cm]{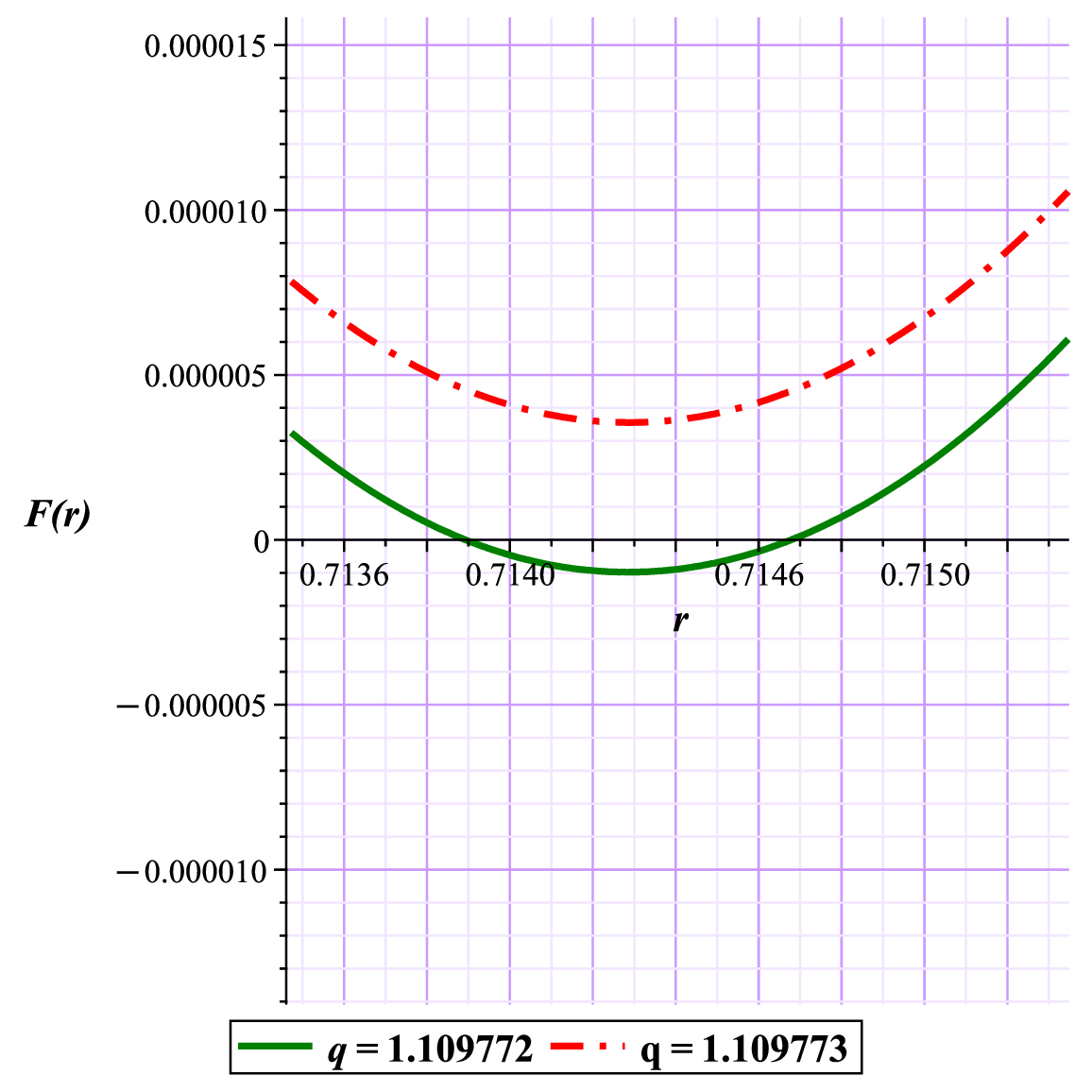}
 \label{11b}}
\subfigure[]{
 \includegraphics[height=5cm,width=5cm]{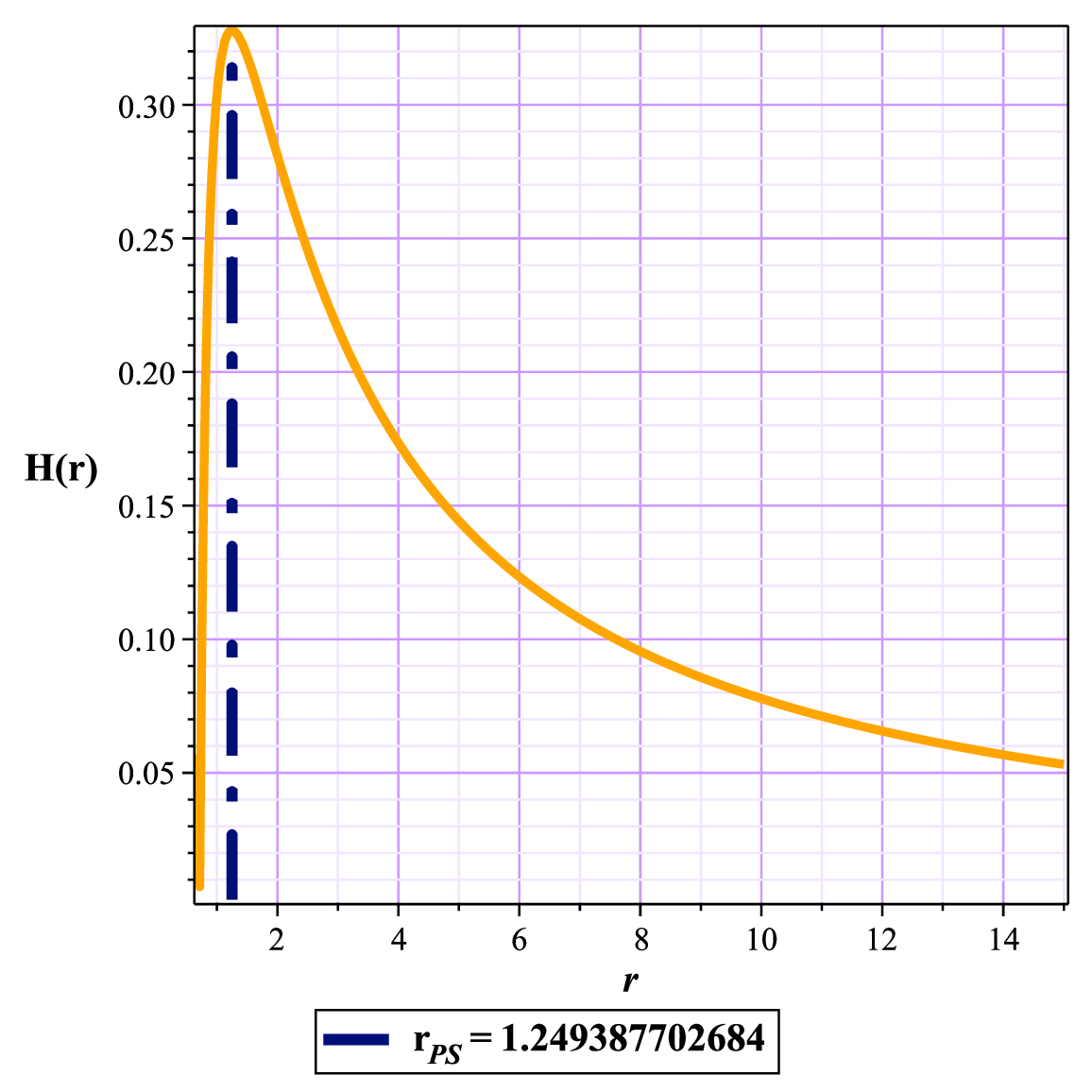}
 \label{11c}}
   \caption{\small{Fig (11a): Metric function with $a = 0.6, \Xi = 10^{-5}, \alpha = -0.1, m = 1$  and different $ q $ for model with Cloud of Strings, Fig (11b): Metric function with $q=1.109772$ and $q=1.109773$ , (11c): The topological potential H(r) with unstable photon sphere for  NCEGB with Cloud  of Strings}}
 \label{11}
\end{center}
\end{figure}

In our previous paper \cite{1}, after defining the initial and common terms in the physics literature for extremal and super-extremal black holes, we addressed the necessity of the existence of super-extremal particles and consequently super-extremal black holes. We stated that, from the perspective of WGC, the existence of both is necessary to maintain the process of evaporation and evolution of a black hole and to prevent the formation of naked singularities. We also mentioned that finding models that meet the necessary conditions to maintain the super-extremal state in black hole form could be a clear and evident step towards fulfilling the WGC condition, and these models could be candidates for examining this conjecture.\\
In this paper, in addition to considering the finding of such models as the main objective, we feel that to achieve more precise results, it is better to expand the initial definitions of extremal and super-extremal black holes.\\
At first glance, for studying the WGC, when the Cauchy horizon coincides with the event horizon and the charge-to-mass ratio equals one, we consider a black hole to be extremal ($q/m = 1$). When the event horizon disappears and the charge-to-mass ratio exceeds one, the structure is considered super-extremal ($q/m > 1$).\cite{31,32,33,34}.\\
These propositions logically consist of two parts:\\
•  $\text{Extremal} = (\text{ The coincides horizons}) \wedge (\text{existence of extremal condition (q/m = 1)})$\\
•  $\text{Super-extremal} = (\text{horizon removal}) \wedge (\text{existence of super-extremal condition (q/m > 1)})$\\
As long as both parts of the proposition are not true, the proposition is considered logically false. This is the situation which exactly we are facing here: models that, despite having super-extremal conditions in terms of charge, still maintain their black hole characteristics with an event horizon and a photon sphere. This phenomenon was observed in the Born-Infeld model \cite{1} and now we can see it in our new cases with $\alpha > 0.1$ and $\alpha < 0$.
For example, for the Charged Gauss-Bonnet black hole with Cloud of Strings model, for the choices $a = 0.6, \Xi = 10^{-5}, \alpha = -0.1, m = 1$, the structure, despite having superextreme conditions in terms of charge, still has the event horizon, its effective potential function is continuous, and has an unstable photon sphere, Fig. (\ref{11}).
Or similarly in the case of $\alpha>0.1$ for the choices $a = 0.6, \Xi = 10^{-7}, \alpha = 0.3, m = 3$,we have Fig. (\ref{12}):
\begin{figure}[H]
 \begin{center}
 \subfigure[]{
 \includegraphics[height=5.5cm,width=6cm]{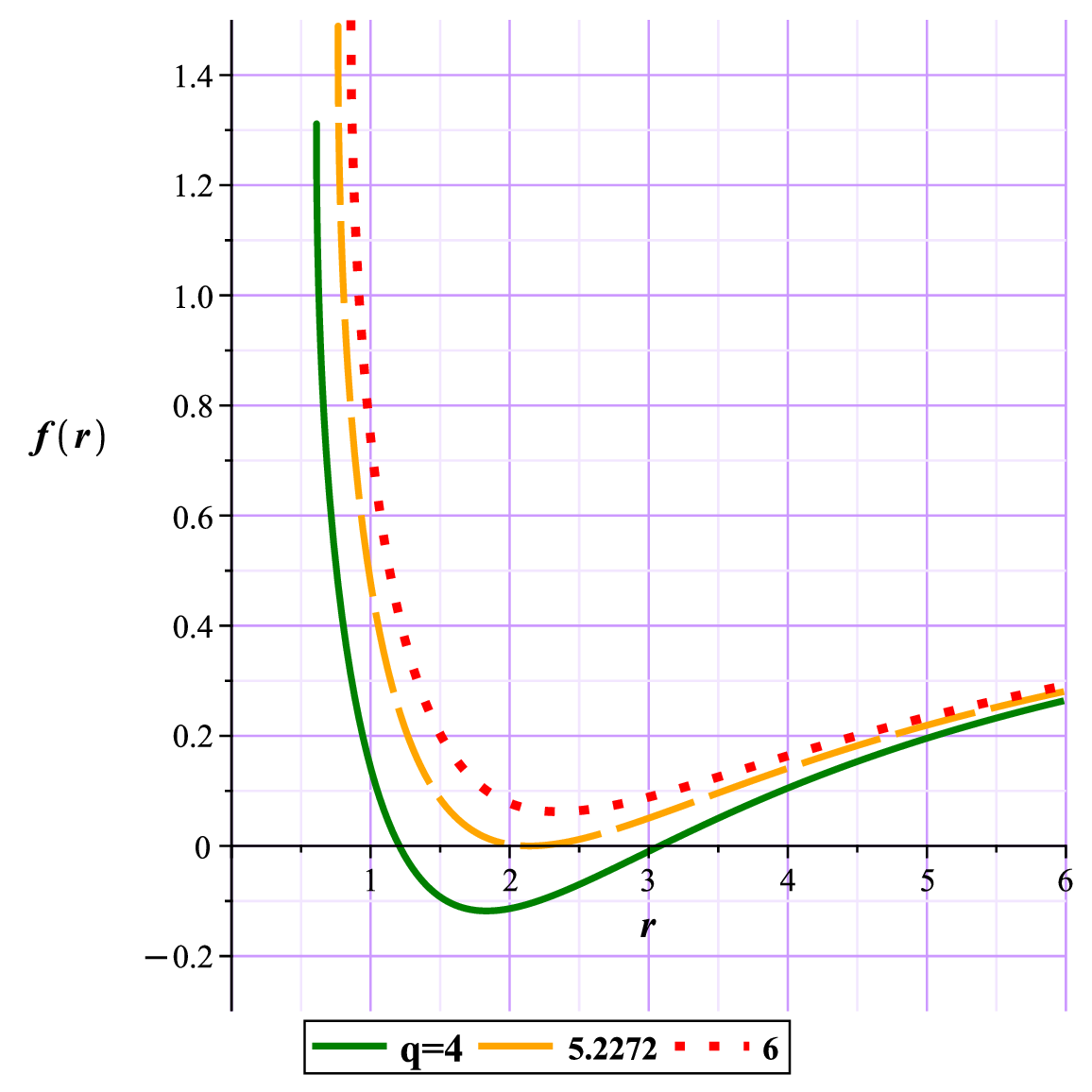}
 \label{12a}}
 \subfigure[]{
 \includegraphics[height=5cm,width=6cm]{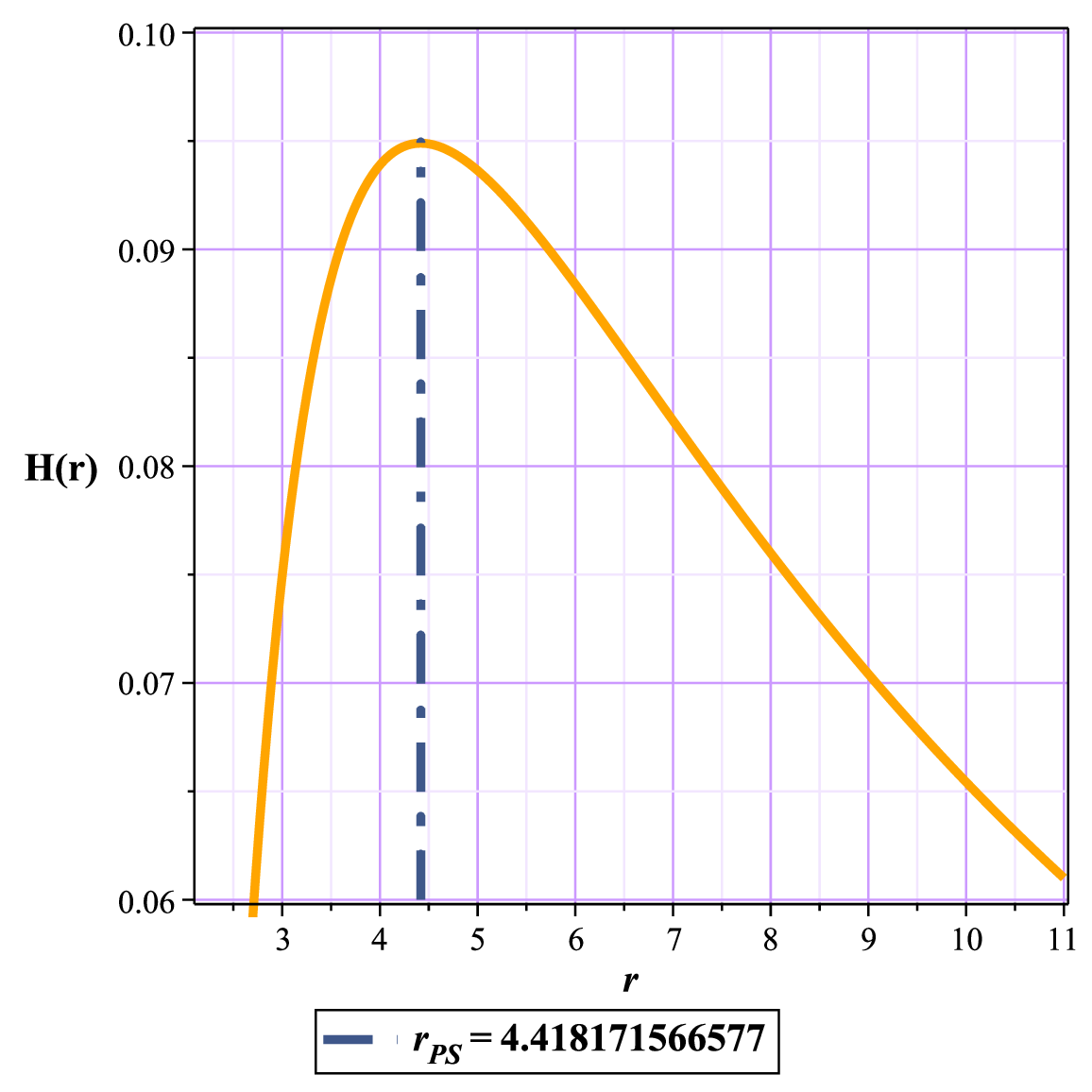}
 \label{12b}}
   \caption{\small{Fig (12a): Metric function with $a = 0.6, \Xi = 10^{-7}, \alpha = 0.3, m = 3$  and different $ q $ for model with Cloud of strings, (12b): The topological potential H(r) with unstable photon sphere for  NCEGB with Cloud of Strings,}}
 \label{12}
\end{center}
\end{figure}
Therefore, on one hand, although these states cannot be considered super-extremal black holes, on the other hand, the increased likelihood of the existence of super-extremal particles in these models can itself be regarded as significant evidence for examining and fulfilling the conditions of the WGC.\\
Before expanding on the main discussion, it would be interesting to point out two points. In most studies on the evidence for the emergence of the Weak Gravity Conjecture (WGC), efforts have been made to find points where the condition of super-extremal charge exists in the presence of an event horizon s \cite{35,36,37,38,39,40,41,42,43,44,45,46,47,48,49}. However, in the models we have examined, a complete band (set of points) appears to satisfy the condition. Additionally, in the case of $\alpha < 0.1$, we observed that there is a critical mass for the model,
which forces us to consider the system to be more massive (or, more precisely, more energetic) compared to other states to maintain a black hole structure. Interestingly, this increase in mass seems to significantly enhance the system's capacity to withstand additional charge, allowing it to become charged up to 1.85 times its mass. This indicates a much broader scope for observing the WGC. It seems that this increase in mass (or energy) acts as a magnifying glass, allowing for a better observation of the landscapes.
\subsection{$ Temperature $ }
\begin{figure}[H]
 \begin{center}
 \subfigure[]{
 \includegraphics[height=5.5cm,width=6cm]{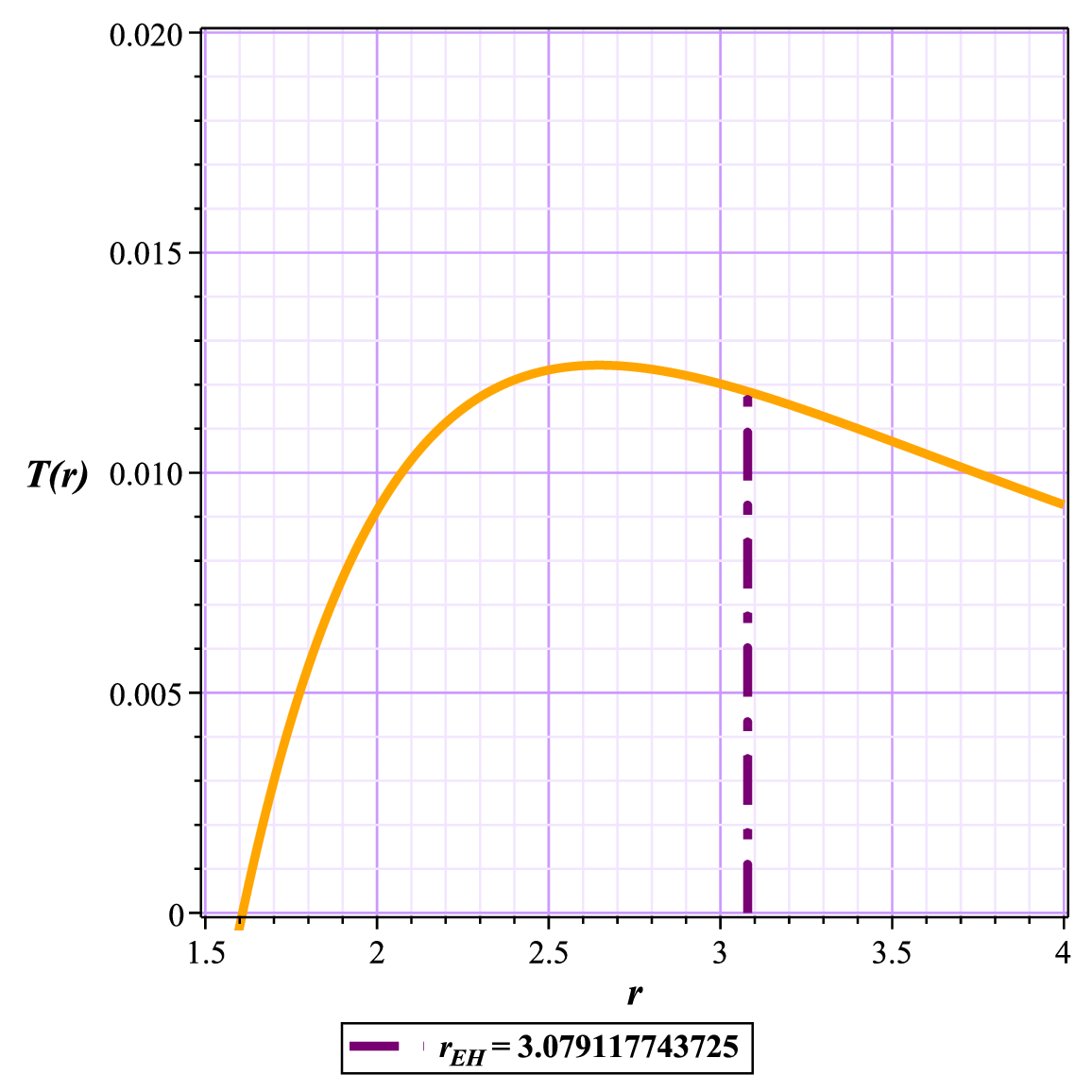}
 \label{13a}}
 \subfigure[]{
 \includegraphics[height=5cm,width=6cm]{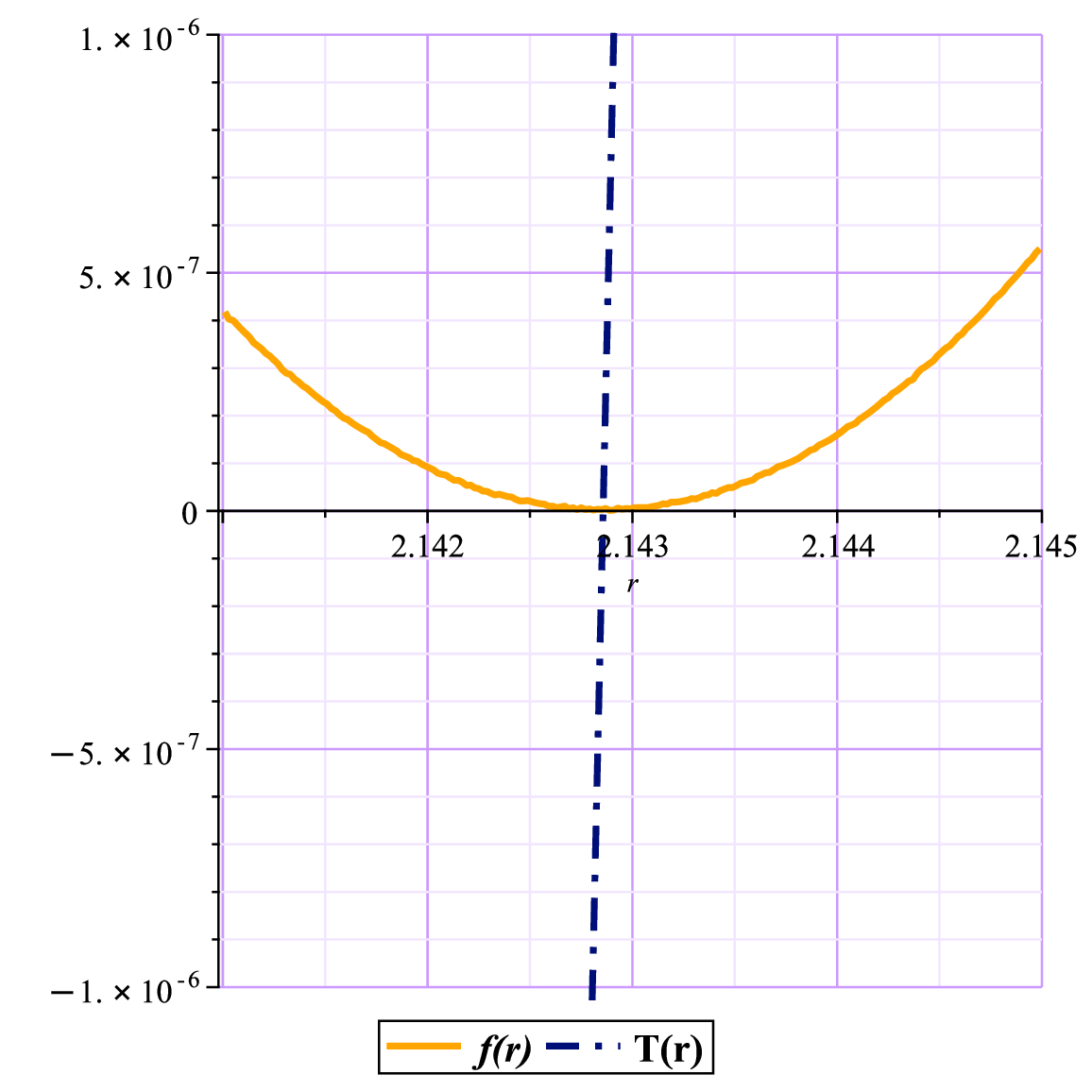}
 \label{13b}}
   \caption{\small{Fig (13a): Temperature function with $a = 0.6,q=4, \Xi = 10^{-7}, \alpha = 0.3, m = 3$  VS radius of Event Horizon, (13b): The confluence of Temperature function and  Metric function for the case $\alpha>0.1$ at the charge tolerance limit q=5.2272174 for  NCEGB with Cloud  of Strings,}}
 \label{13}
\end{center}
\end{figure}

As we can see in Fig. (\ref{11}), Fig. (\ref{12}), it appears that there is an issue with the provided definition of an extremal black hole and the super-extremal form, as there are clearly states that are super-extremal in terms of charges but still possess the event horizon and the photon sphere. It seems more appropriate to refine the definition and convert it into a three-propositions form. For this purpose, we utilize the concept of temperature \cite{49.1,52}.
We know that when the structure is in the extremal form, the Hawking radiation of extremal black holes is considered non-thermal, it means temperature of the event horizon approaches zero, which means the cessation of radiation through energy \cite{50}. However, our studies show that when our structure is super-extremal in terms of charge, the event horizons have a non-zero temperature Fig. (\ref{13a}). Yet, when we reach the charge tolerance limit, where the Cauchy horizon and the event horizon coincide, the temperature drops to zero, Fig. (\ref{13b}).
This seems to be exactly where the structure has reached its extreme. So in the WGC study, the definition of an extremal black hole can be written as follows:\\\\
\text{Extremal Black Hole} = (\text{The coincides  Horizon}) $\wedge (\text T(r_{H})=0) \wedge $ \text{existence extremal or super-extremal condition $(q/m \geq 1)$}.\\\\
Any state beyond the above definition would be a superextreme black hole.\\ Now, considering these concepts, let's return to the WGC. When a black hole reaches the extremal form as defined above, it loses its thermal radiation, meaning it is not expected to emit significant amounts of Hawking radiation. Therefore, at first glance, it appears "static" in terms of energy exchange. However, due to the highly fragile nature of this state in terms of stability and the surrounding environmental conditions, these black holes cannot maintain this state indefinitely. Now what scenarios could be ahead?\\
\begin{figure}[H]
 \begin{center}
 \subfigure[]{
 \includegraphics[height=5.5cm,width=6cm]{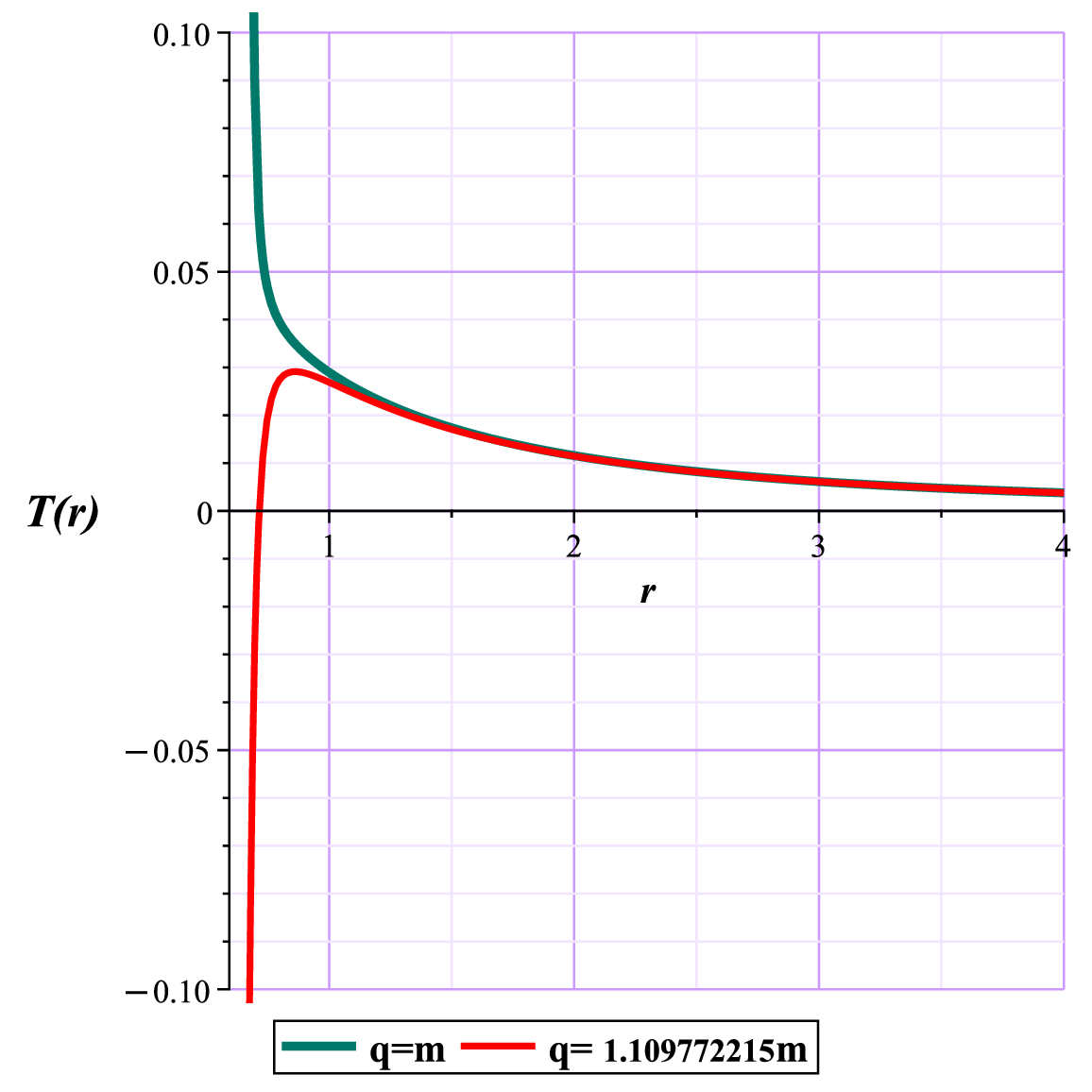}
 \label{14a}}
 \subfigure[]{
 \includegraphics[height=5cm,width=6cm]{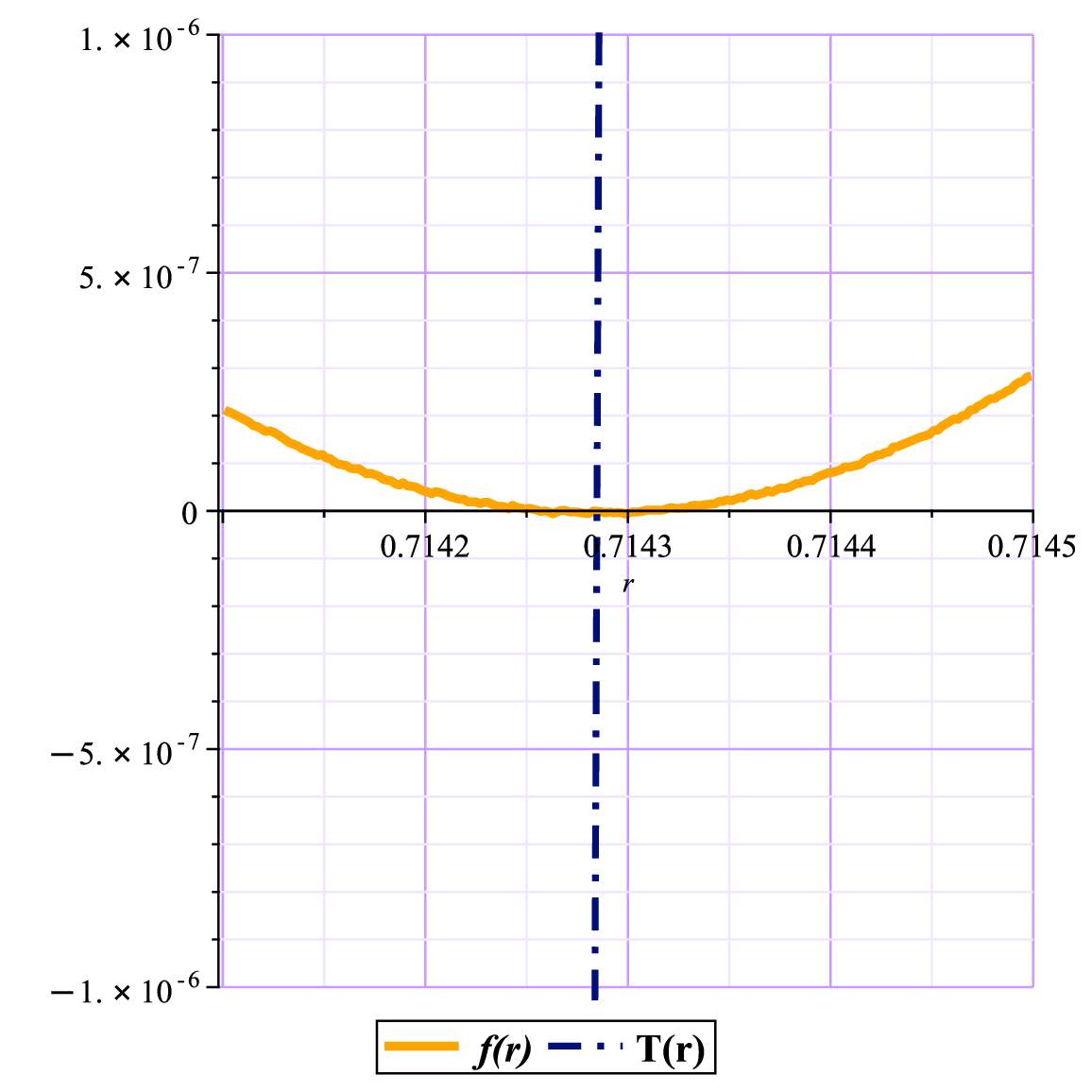}
 \label{14b}}
   \caption{\small{Fig (14a): Temperature function with $a = 0.6,\Xi = 10^{-7}, \alpha =- 0.1, m = 1$  for $ q = m $ and $ q = 1.109772215 m $, (14b): The confluence of Temperature function and  Metric function for the case $\alpha<0$ at the charge tolerance limit q=1.109772215 for  NCEGB with Cloud  of Strings,}}
 \label{14}
\end{center}
\end{figure}
In Fig. (\ref{14a}), we plotted the temperature graph in two states: $ q = m $ and $ q = 1.109772215 m $, which is the charge tolerance limit for a NC black hole with $ \alpha < 0$. As we can see, the green curve corresponds to the sub-extremal black hole, and the red curve corresponds to the extremal black hole with the new definition. When the black hole approaches the extremal state, the behavior of the temperature function with respect to the radius completely changes. In the green curve, the temperature is entirely in the positive region, following the causal world of general relativity, as positive temperature always has a physical interpretation.
However, as it approaches the extremal state, the temperature graph starts from negative temperatures.\\
Now for explanation, we can either take the harder path and assume that in this state, the black hole slips into an unknown dynamic that can withstand and interpret negative temperatures, which means that we are effectively losing our black hole into the swamp plan.
Or we can take the simpler path, which is the path of censorship.
In this case, we assume that, instead of having negative temperatures or naked singularities, with the increase in super-extreme particles and the strengthening of electromagnetism, the necessary conditions are practically provided for mechanisms that can lead to the escape of charged particles. Processes such as Schwinger pair production may occur in the strong field around an extremal black hole, creating pairs of virtual particles. Through particle creation, it is possible for the black hole to lose charge, meaning the black hole returns to sub-extreme and normal conditions.
In this interpretation, by utilizing conditions where electromagnetic fields temporarily dominate over gravity, we can maintain the concepts of black holes, event horizons, and, in short, WCCC and all known physics without concerns about negative temperatures, naked singularities, instabilities, or any other non-causal relativistic issues. Therefore, in reality, WGC acts as a powerful protector for the WCCC.\\
Thereupon, it seems that based on the concepts discussed, we have identified three states in the Born-Infeld and Charged String Cloud models that, in addition to being super-extremal in terms of charge, also meet the more conditions necessary to satisfy the WGC.
\section{Conclusions}
In the previous article \cite{1}, we explained that NC black holes, instead of considering a point singularity, attempt to eliminate the central singularity using a Gaussian distribution instead of the Dirac delta function. This NC parameter, $\Xi$, appears like a sheath around the singularity, allowing us to extend our geometry as close to the center as possible. Clearly, the thinner this sheath, the greater the accuracy of our work. Therefore, from a less precise perspective, one might consider this parameter, which is not dependent on the physical structure of the model but rather has a geometric aspect, as a criterion for matching the model's accuracy with physical reality.
Based on these concepts, we aimed to study the photon sphere and time-like orbits for this class of black holes to examine the impact of this change in mass structure on these geodesics. Also, based on previous studies\cite{19,20,21,51,51.1}, we attempted to determine the effective range of the model parameters concerning the photon sphere and ultimately soughed models that satisfy the conditions for the WGC. Accordingly, we first examined the Schwarzschild-like model (Apendix A), then the model influenced by linear charge (Reissner-Nordström-like), and then the model influenced by the non-linear Born-Infeld field \cite{1}. Afterward we  moved on to the models that primarily affect gravity and curvature, namely the Gauss-Bonnet model, and finally examined a combined model of the linear field and Gauss-Bonnet, which is enveloped by a string cloud to observe the maximum combined effect.
\begin{center}
\begin{table}[H]
  \centering
\begin{tabular}{|p{3cm}|p{3cm}|p{1.1cm}|p{3cm}|p{5cm}|}
  \hline
  \centering{Black hole type}  & \centering{Fix parametes} &\centering{*Eff-Pa}& Black hole Renges &\ $Description$\\[3mm]
   \hline
  \centering{N-C SCHWARZSCHILD} & \centering $ m = 1  $ & \centering{$\Xi $} & \centering $0<\Xi \leq 0.27581 $&\ Appendix A\\[3mm]
   \hline
 \centering{N-C in 4D E-G-B} & \centering $  \alpha = 0.39, m = 1 $ & \centering{$\Xi,\alpha>0.5$} &\centering $0<\Xi \leq 0.1081 $&\ There is a critical $\alpha$ for the model (0.5) beyond which the model has only singularity form. \\[3mm]
   \hline
   \centering{Charged NCBH } & \centering $ q = 0.6, m = 1 $ & \centering{$\Xi$} & \centering $ 0<\Xi \leq 0.2184  $ &\ The model lacks a superextreme charging form \cite{1}. \\[3mm]
   \hline
   \centering{NC B-I($\Lambda<0$) } & \centering $  b=0.5,\Lambda=-1,q = 0.6, m = 1 $ & \centering{$\Xi$} & \centering $ 0<\Xi \leq 0.0611   $ &\ The model have a superextreme charging form.It can also appear in the form of an extremal black hole (WGC candidate) \cite{1}. \\[3mm]
   \hline
   \centering{NC-CLOUD STRINGS BH } & \centering $   a = 0.6, q = 0.6,\alpha = -0.1, m = 1 $ & \centering{$\Xi,\alpha<0 $} & \centering $ 0<\Xi \leq 0.01462   $ &\ The model have a superextreme charging form.It can also appear in the form of an extremal black hole (WGC candidate). \\[3mm]
   \hline
   \centering{NC-CLOUD STRINGS BH } & \centering $  a = 0.8, q = 0.6, \alpha = 0.01, m = 1 $ & \centering{$\Xi,0<\alpha<0.1$} & \centering $ 0<\Xi \leq 0.00186   $ &\ The model lacks a superextreme charging form. \\[3mm]
   \hline
   \centering{NC-CLOUD STRINGS BH } & \centering $  a = 0.6, q = 1, \alpha = 0.3, m = 3 $ & \centering{$\Xi,\alpha>0.1$} & \centering $ 0<\Xi \leq 0.1074 $ &\ The model have a superextreme charging form.It can also appear in the form of an extremal black hole (WGC candidate). \\[3mm]
   \hline
   \end{tabular}
   \caption{*Eff-Pa: Effective Parameters-The parameter whose changes have been most considered.}\label{5}
\end{table}
 \end{center}
As shown in Table (VI), the largest range of $\Xi $ (the least accurate) corresponds to the Schwarzschild-like model, which was expected. Adding a linear charge, although it has an impact on the range, this effect is clearly not very significant. However, the addition of a non-linear field to the model seems to result in considerable accuracy.
A very interesting point in comparing the charged model with the Born-Infeld model is that the addition of the field does not seem to affect the effective gravitational influence range, as for both models, the maximum gravitational effect (black hole + naked singularity) remains within the range of$ 0 < \Xi < 0.3273$. However, the spatial distribution between these two states within this range clearly indicates the higher accuracy of the Born-Infeld model \cite{1}.\\
Although it is often stated that studying curvature through Gauss-Bonnet in higher dimensions shows the most significant impact on the results, and practically in four dimensions, a significant effect cannot be observed, the comparison related to the Gauss-Bonnet-like model with the Schwarzschild-like model shows a relatively significant difference in the permissible range. Not only the black hole range, but also the effective gravitational influence range, has a significant difference. In the Gauss-Bonnet-like state, the model seems to have even more considerable accuracy in matching with physics and reality compared to the charged model (Table 1, \cite{1}). However, it seems that the best results pertain to the final combined model, which includes not only the linear charge and Gauss-Bonnet parameter but also the influence of the string cloud. The last case in the above table cannot be included in this comparison due to mass manipulation, where to have more conventional choices for the Gauss-Bonnet parameter and to maintain the constraints of the string cloud parameter, we had to consider the black hole to be more massive (more energetic) because for this state, there was a critical mass below which we faced issues.\\
In the study of TCOs, all models not only confirmed the standard presented in \cite{22} but also extended its comprehensiveness, demonstrating the spatial classification process over larger ranges. However, perhaps the most interesting point in the study of TCOs was the emergence of a forbidden region near the center in the string cloud model, which was not observed in \cite{22} or other studied models, as shown in Fig. (\ref{10}) and Fig. (\ref{9b})).\\
The core objective of this article was to search for models that could meet the necessary conditions for studying the WGC. In previous work \cite{1}, we argued that the existence of super-extremal particles plays a crucial role in maintaining the black hole evaporation process, And the models that could be represented these superextreme particles in black hole form could play a key role in studying the WGC \cite{1}. For Instance, we demonstrated in that work that the Born-Infeld model possesses this capacity. However, is the presence of such charges alone sufficient?
In this article, we explained that there are other conditions to consider when defining extremal black holes in the context of WGC. Also, we explained that how WGC can support and protect the Weak Cosmic Censorship Conjecture and our physical and causal knowledge.
We also showed that instead of single points, there is a range of points where both the Born-Infeld and string cloud models can accommodate super-extremal charges while maintaining their black hole form. This means that, in addition to having an event horizon and a photon sphere, they also meet the defined conditions for extremality. Therefore, we suggest that these models can be considered as evidence and candidates for WGC study.
\section{Appendix A: NON-COMMUTATIVE SCHWARZSCHILD BLACK HOLE }
The metric for the black hole is \cite{52}:
\begin{equation}\label{(15)}
f =1-\frac{4 \gamma \! \left(\frac{3}{2},\frac{r^{2}}{4 \Xi}\right) m}{\sqrt{\pi}\, r}.
\end{equation}
With respect to Eq. (\ref{(4)}), Eq. (\ref{(5)}) we have:
\begin{equation}\label{(19)}
\begin{split}
H =\frac{\sqrt{1-\frac{4 \left(\frac{\sqrt{\pi}}{2}-\frac{\sqrt{\frac{r^{2}}{\Xi}}\, {\mathrm e}^{-\frac{r^{2}}{4 \Xi}}}{2}-\frac{\sqrt{\pi}\, \mathrm{erfc}\left(\frac{\sqrt{\frac{r^{2}}{\Xi}}}{2}\right)}{2}\right) m}{\sqrt{\pi}\, r}}}{\sin \! \left(\theta \right) r}.
\end{split}
\end{equation}
\begin{equation}\label{(20)}
\begin{split}
\phi^{r}=\frac{3 \csc \! \left(\theta \right) \left(\left(\mathrm{erf}\! \left(\frac{\sqrt{\frac{r^{2}}{\Xi}}}{2}\right)-\frac{r}{3}\right) \Xi  \sqrt{\pi}-\frac{\sqrt{\frac{r^{2}}{\Xi}}\, \left(r^{2}+6 \Xi \right) {\mathrm e}^{-\frac{r^{2}}{4 \Xi}}}{6}\right)}{\sqrt{\pi}\, \Xi  r^{3}}.
\end{split}
\end{equation}
\begin{equation}\label{(21)}
\begin{split}
\phi^{\theta}=-\frac{\sqrt{1-\frac{4 \left(\frac{\sqrt{\pi}}{2}-\frac{\sqrt{\frac{r^{2}}{\Xi}}\, {\mathrm e}^{-\frac{r^{2}}{4 \Xi}}}{2}-\frac{\sqrt{\pi}\, \mathrm{erfc}\left(\frac{\sqrt{\frac{r^{2}}{\Xi}}}{2}\right)}{2}\right) m}{\sqrt{\pi}\, r}}\, \cos \! \left(\theta \right)}{\sin \! \left(\theta \right)^{2} r^{2}}.
\end{split}
\end{equation}
\begin{center}
\begin{table}[H]
  \centering
\begin{tabular}{|p{3cm}|p{4cm}|p{5cm}|p{1.5cm}|p{2cm}|}
  \hline
  \centering{NCSCH BH}  & \centering{Fix parametes} &\centering{Conditions}& *TTC&\ $(R_{PLPS})$\\[3mm]
   \hline
  \centering{unstable photon sphere} & \centering $  m = 1  $ & \centering{$0< \Xi \leq 0.27581 $} & $-1$&\ $2.980321831$\\[3mm]
   \hline
 \centering{naked singularity} & \centering $ m = 1 $ & \centering{$0.27581< \Xi \leq 0.4006 $} &\centering $0$&\ $-$ \\[3mm]
   \hline
   \centering{*Unauthorized area} & \centering $ m = 1 $ & \centering{$ \Xi > 0.4006 $} & \centering $ nothing $ &\ $-$ \\[3mm]
   \hline
   \end{tabular}
   \caption{*Unauthorized region: The region with negative or imaginary roots of $\varphi$.\\ $R_{PLPS}$: the minimum or maximum possible radius for the appearance of an unstable photon sphere.}\label{1}
\end{table}
 \end{center}
 \section{Appendix B: NON-COMMUTATIVE 4D Einstein-Gauss-Bonnet }
 \begin{center}
\textbf{$\alpha = 0.05$ }
\end{center}
\begin{center}
\begin{table}[H]
  \centering
\begin{tabular}{|p{3cm}|p{4cm}|p{5cm}|p{1.5cm}|p{2cm}|}
  \hline
  \centering{NCEGB BH}  & \centering{Fix parametes} &\centering{Conditions}& *TTC&\ $(R_{PLPS})$\\[3mm]
   \hline
  \centering{unstable photon sphere} & \centering $ \alpha = 0.05, m = 1  $ & \centering{$0< \Xi \leq 0.2541 $} & $-1$&\ $2.940731529$\\[3mm]
   \hline
 \centering{naked singularity} & \centering $\alpha = 0.05, m = 1 $ & \centering{$0.2541< \Xi \leq 0.3822 $} &\centering $0$&\ $-$ \\[3mm]
   \hline
   \centering{*Unauthorized area} & \centering $\alpha = 0.05, m = 1 $ & \centering{$ \Xi > 0.3822 $} & \centering $ nothing $ &\ $-$ \\[3mm]
   \hline
   \end{tabular}
   \caption{*Unauthorized region: The region with negative or imaginary roots of $\varphi$.\\ $R_{PLPS}$: the minimum or maximum possible radius for the appearance of an unstable photon sphere.}\label{1}
\end{table}
 \end{center}

\end{document}